\documentclass[a4]{article}
\usepackage[left=2cm,right=2cm,top=2cm,bottom=2cm,bindingoffset=0cm]{geometry}
\usepackage{multirow}

\usepackage{amsfonts, amsmath, amssymb}
\usepackage{graphicx, float}
\usepackage{subcaption}

\newcommand{\be}{\begin{equation}}
\newcommand{\ee}{\end{equation}}

\graphicspath{{./pictures/}}

\begin{document}
	\title{ON DIRECT SEARCH FOR DARK MATTER IN SCATTERING PROCESSES WITHIN YUKAWA MODEL}
	\author{V.V.~Skalozub$^1$, M.S.~Dmytriiev$^2$, \\ $^1$Skalozubv@daad-alumni.de, $^2$dmytrijev\_m@ffeks.dnu.edu.ua, \\
	Oles Honchar Dnipro National University, \\
    72, Gagarin Ave., Dnipro 49010, Ukraine}

    \maketitle

    \begin{abstract}
    	Nowadays, no dark matter candidates have been discovered. We consider two possible reasons for that, both related to the approach of on-peak resonance searching for. As is believed usually, a new particle suits the conditions that the ratio of the width to the mass is less than $1-3 \%$ and a narrow width approximation (NWA) is applicable to identify such type resonant peak in the invariant mass spectrum of the collision products.
    	
    	In the present paper, in the framework of generalized Yukawa model, we find out the properties of the searched particle when its width is larger than a maximal one expected during experiments and so this state could be confused with a noise. We also ascertain the values of particle's parameters when the NWA is not applicable and estimate the  width value when it happens. These estimations are relevant to interactions between the Standard model and dark matter particles. Such approach is focused on the role of couplings and mass values introduced in the model describing interaction of visible and dark matter.
    \end{abstract}


    \section{Introduction}

    Nowadays, no particles-candidates for dark matter (DM)  have been found, yet. There is  a long list of such particles entering different  models relevant to various energy scales. In what follows, we consider why it could be so within the standard treating of scattering experiment data. It is usually assumed that the resonances of the unknown particles are narrow ones and their typical width $\Gamma$ is about $1-3\%$ of the peak's mass $M$, so $\Gamma \ll M$. This assumption allows to apply the NWA to detect  the resonances in the total cross-section. In this approach, the interference between visible and dark particles could be neglected in the total cross section. At the same time, there are many models beyond the Standard model (SM) present in the literature, which anticipate new particles to have wide resonances. Consequently, such signals can be confused with a noise if they appear in experimental data. In present paper, we analyze how the widths of new physics peaks depend on masses and couplings introduced in some underlying model of the DM.

    We carry out our investigation in the framework of the generalized Yukawa model. DM is represented there as the heavy Dirac fermion $\Psi$ and scalar $\chi$ fields. Visible matter is described as light scalar field $\phi$ and the doublet of light fermion fields $\psi_1$ and $\psi_2$. The latter two interact with bosons $\phi$ and $\chi$ through different Yukawa's couplings. Using this model, we obtain the width of dark boson $\chi$ for certain values of the model parameters. Such approach is different from that used in some non-NWA peak investigations presented in, for instance, \cite{bibl:jung, bibl:moretti-nwa, bibl:moretti-dark-nwa}. In these papers the width of the new particle is adopted to be an  arbitrary free parameter.

    In the literature, there exist numerous models of DM  such as SUSY particles (neutralinos) \cite{bibl:drees}, neutral $Z^{\prime}$ bosons \cite{bibl:leike, bibl:langacker, bibl:gulov}, sterile neutrinos \cite{bibl:boyarsky}, etc.  In these models, it is assumed that DM candidate has a certain  group of symmetry or specified  couplings to  other fields. On the contrary,  we consider all the parameters of DM as free, and do not  limit our treatment by  a certain gauge group. Also, in what follows  we do not take into consideration the astrophysical limits on the couplings between visible and dark sectors. These couplings  affect the production and decay rates of the DM, and so its relic abundance in the Universe.
    Such problems could be investigated in the context of energy distribution of the dark  and visible particles.  All these  topics are left beyond the scope of the present paper.

    Below we are concentrating  on the analysis  of the particle parameter values for which  the dark resonance does not satisfy the NWA bounds, and becomes invisible in the direct search for it. Then we derive the  necessary constraints on the  properties of the DM sector. Finally, we build cross-section of a chosen scattering process mediated by dark particles, and investigate interference and resonant contributions of the latter to it.

    The paper is organized as follows. In next section we introduce our model and discuss the mixing of scalar fields, which appears at one-loop level. We also  define  the corresponding mixing angle. In sect. 3, we consider width of the dark particle resonance in the context of the chosen s-channel scattering process. Then we estimate the values of the parameters when this width exceeds the NWA limits, and provide limitations for the mixing angle. In sect. 4 we investigate contribution of the dark particles to the interference and resonant terms in the process cross-section. We summarize  and discuss our  results  in the context of comparisons with a number of DM models in the last section.

    \section{The model}

\begin{figure}[h]
    	\vskip1mm
    	\begin{center}
    		\begin{tabular}{ccc}
    			\includegraphics[scale=0.4]{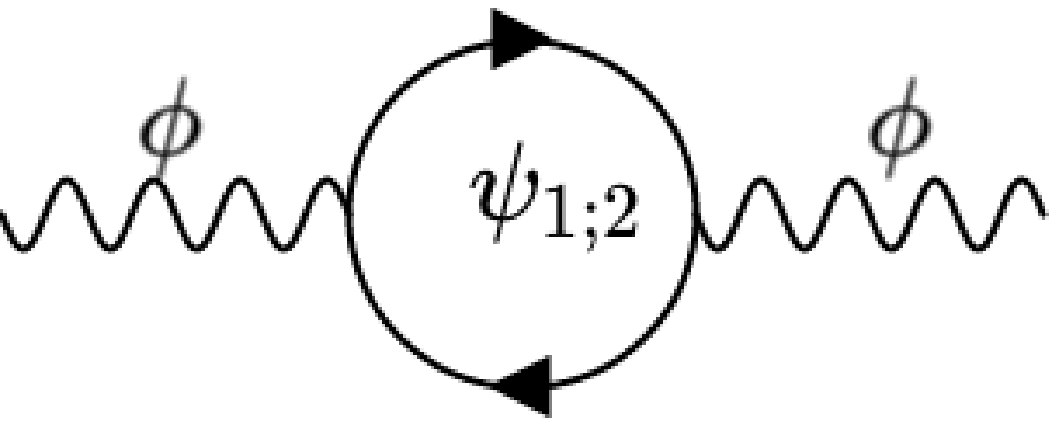} & \includegraphics[scale=0.4]{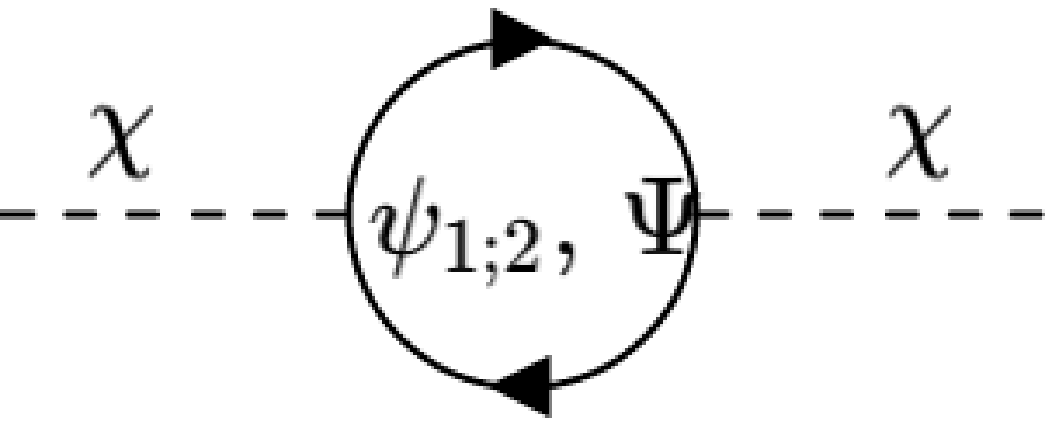} &
    			\includegraphics[scale=0.4]{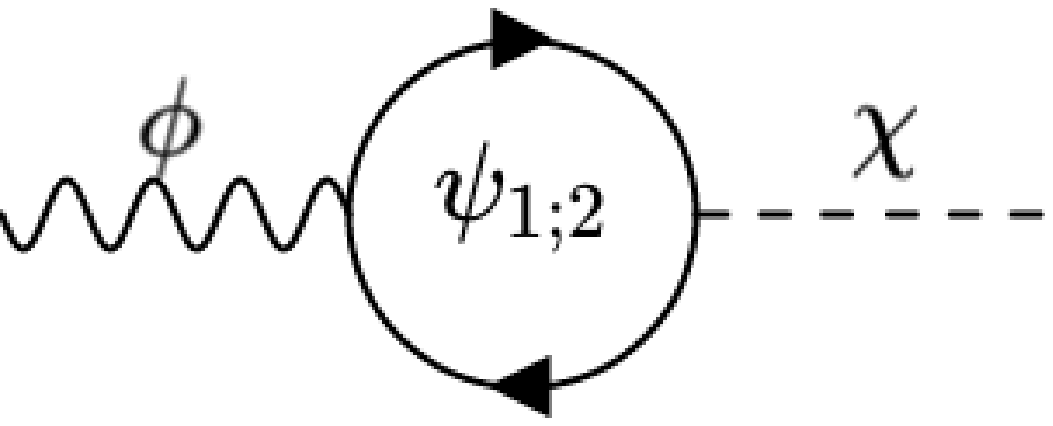}
    		\end{tabular}
    	\end{center}
    	\vskip-3mm
    	\caption{Fermion loop corrections to the two-point Green functions $\left<0|T\phi(x_1)\phi(x_2)|0\right>$, $\left<0|T\chi(x_1)\chi(x_2)|0\right>$ and $\left<0|T\phi(x_1)\chi(x_2)|0\right>$, respectively. Fermion fields are listed inside loops  which they contribute to.}
    	\label{fig:one-loop-mixing}
    \end{figure}

    We start with the Lagrangian
    \begin{equation}
    \label{lagrangian}
    \begin{split}
    & \mathcal{L} = \frac{1}{2}\left[\left(\partial_{\mu}\phi\right)^{2}-\mu^{2}\phi^{2}\right]+\frac{1}{2}\left[\left(\partial_{\mu}\chi\right)^{2}-\Lambda^{2}\chi^{2}\right] - \lambda\phi^{4}-\rho\phi^{2}\chi^{2}-\xi\chi^{4} + \\
    & + \sum\limits_{a = 1;2}\bar{\psi}_a\left(i\gamma^{\mu}\partial_{\mu} - g_{\phi}\phi - g_{\chi}\chi-m_a\right)\psi_a + \bar{\Psi}\left(i\gamma^{\mu}\partial_{\mu}-M - G_{\chi}\chi\right)\Psi.
    \end{split}
    \end{equation}
    Dark fermions $\Psi$, having only coupling $G_{\chi}$ to scalars in the dark sector,  do not interact with visible bosons $\phi$. Due to the presence of both couplings $g_{\phi}$ and $g_{\chi}$  the scalar fields are mixed at the one-loop level, with the mixing angle $\theta_{mix}$. Value of $\theta_{mix}$ regulates intensity of interaction between visible and dark sectors. In particular, this mixing expresses itself as the non-diagonal loop corrections coming  from fermionic loops in the scalar two-point Green functions, depicted in fig.\,\ref{fig:one-loop-mixing}.
    We introduce the mixing angle as $O(2)$-rotation, which diagonalizes the mass matrix of  scalar fields at zero momentum. This matrix emerges from the effective potential of bosons for constant fields $\phi$ and $\chi$. The total effective potential $V_{eff}$ consists of the tree-level part $V_{eff}^{(tree)}$ and correction, which goes from Yukawa's interaction, obtained after integration over virtual fermions,
    \begin{equation}
    \label{nonRenormPoten}
    \begin{split}
    & V_{eff}^{(tree)} = \frac{1}{2}\mu^2\phi^2 + \frac{1}{2}\Lambda^2\chi^2 + \lambda\phi^4 + \rho\phi^2\chi^2 + \xi\chi^4 \\
    & V_{eff} = V_{eff}^{(tree)} + \lim\limits_{\epsilon\rightarrow 0}\frac{\kappa^{2\epsilon}}{8\pi^2}\int\limits_0^{\infty}\frac{ds}{s^{3 - \epsilon}}\left\{\sum_{a=1;2}e^{-s(m_a + g_{\phi}\phi + g_{\chi}\chi)^2} + e^{-s(M + G_{\chi}\chi)^2}\right\} = \\
    & = V_{eff}^{(tree)} - \frac{1}{16\pi^2}\left\{\sum\limits_{a = 1;2}\left(m_a + g_{\phi}\phi + g_{\chi}\chi\right)^4\ln\frac{\left(m_a + g_{\phi}\phi + g_{\chi}\chi\right)^2}{\kappa^2} + \right. \\
    & + \left. \left(M + G_{\chi}\chi\right)^4\ln\frac{\left(M + G_{\chi}\chi\right)^2}{\kappa^2}\right\} + \frac{\Delta}{16\pi^2}\left\{\sum\limits_{a = 1;2}\left(m_a + g_{\phi}\phi + g_{\chi}\chi\right)^4 + \left(M + G_{\chi}\chi\right)^4\right\},\qquad \\
    & \Delta = \frac{1}{\epsilon} - \gamma_E\quad (\epsilon\rightarrow 0).
    \end{split}
    \end{equation}
    Here $\gamma_E$ is the Euler-Mascheroni number, $\Delta$ is divergent constant, and $\kappa$ is  a renormalization point. We adopt $\kappa$ arbitrarily and take into consideration the  effects of the model couplings and masses, only. Integral in \eqref{nonRenormPoten} accounts for fermionic corrections to the bosonic masses and self-interaction constants $\lambda$, $\rho$ and $\xi$. The components of the bosonic mass matrix $M_{ab}^2$ are given as the coefficients of the $V_{eff}$ Taylor's series expansion calculated  at $\phi_0$ and $\chi_0$:
    \begin{equation}
    \label{massMatrixDef}
    \begin{split}
    & M_{11}^2 = \frac{\partial^2V_{eff}}{\partial\phi^2}\Bigr|_{\phi_0;\chi_0},\quad M_{12}^2 = \frac{\partial^2V_{eff}}{\partial\phi\partial\chi}\Bigr|_{\phi_0;\chi_0}, M_{22}^2 = \frac{\partial^2V_{eff}}{\partial\chi^2}\Bigr|_{\phi_0;\chi_0}.
    \end{split}
    \end{equation}
    In what follows, we choose $\phi_0 = 0$ and $\chi_0 = 0$ that sets the renormalization point of bosonic masses. Hence, the mass matrix components are
    \begin{equation}
    \label{fullMassMatrix}
    \begin{split}
    &M_{11}^2 = \mu^2 - \frac{3g_{\phi}^2}{4\pi^2}F + \frac{3g_{\phi}^2}{4\pi^2}\left(m_1^2 + m_2^2\right)\Delta, \\
    &M_{12}^2 = - \frac{3g_{\phi}g_{\chi}}{4\pi^2}F + \frac{3g_{\phi}g_{\chi}}{4\pi^2}\left(m_1^2 + m_2^2\right)\Delta, \\
    &M_{22}^2 = \Lambda^2 - \frac{3}{4\pi^2}\left(g_{\chi}^2 F + G_{\chi}^2 M^2 \ln\frac{M^2}{\kappa^2}\right) + \frac{3}{4\pi^2}\left[g_{\chi}^2\left(m_1^2 + m_2^2\right) + G_{\chi}^2 M^2\right]\Delta, \\
    &F = m_1^2\ln\frac{m_1^2}{\kappa^2} + m_2^2\ln\frac{m_2^2}{\kappa^2}.
    \end{split}
    \end{equation}
    In these equalities we do not account for the terms coming from scalar self-interaction. This is because such terms are proportional to $\phi_0^2$, $\phi_0\chi_0$ or $\chi_0^2$. But since $\phi_0$ and $\chi_0$ are zeroes, their contributions vanish. The mass matrix components,  accounting for the fermionic loop corrections, become divergent and a mass renormalization is required. To do this, we add to the $V_{eff}$ such counter-terms that divergent parts proportional to $\Delta$ in \eqref{fullMassMatrix} vanish\footnote{To perform such renormalization, we use counter-terms $-\frac{1}{2}\delta\mu^2\phi^{(ren)2}$ and $-\frac{1}{2}\delta\Lambda\chi^{(ren)2}$, where $\phi^{(ren)}$ and $\chi^{(ren)}$ are fields rotated on angle $\theta_{ren}$ with respect to $\phi$ and $\chi$. Hence, renormalization procedure in our model consists of $O(2)$-rotation and subtraction of divergent terms. Angle $\theta_{ren}$ is defined as follows:
    	\[\tan{2\theta_{ren}} = \frac{2g_{\phi}g_{\chi}}{g_{\phi}^2 - g_{\chi}^2 - G_{\chi}^2\frac{M^2}{m_1^2 + m_2^2}}\]}.

    Having angle $\theta_{mix}$, we rotate  the scalar fields $\phi$ and $\chi$, turning them to the  basis of "physical" states $\phi^{\prime}$ and $\chi^{\prime}$:
    \[\begin{pmatrix}
    \phi \\ \chi
    \end{pmatrix} = \begin{pmatrix}
    \cos{\theta_{mix}} & -\sin{\theta_{mix}} \\ \sin{\theta_{mix}} & \cos{\theta_{mix}}
    \end{pmatrix} \begin{pmatrix}
    \phi^{\prime} \\ \chi^{\prime}
    \end{pmatrix}.\]
    We pick $\theta_{mix}$ requiring to vanish the non-diagonal term $M_{12}^2$ of the mass matrix for rotated fields $\phi^{\prime}$ and $\chi^{\prime}$. It happens when $\theta_{mix}$ is equal to
    \begin{equation}
    \label{mixingAngleDef}
    \begin{split}
    & \tan{2\theta_{mix}} = \frac{2M_{12}^2}{M_{11}^2 - M_{22}^2} = 2g_{\phi}g_{\chi}F\left[\frac{4\pi^2}{3}\left(\Lambda^2 - \mu^2\right) + \left(g_{\phi}^2 - g_{\chi}^2\right)F - G_{\chi}^2 M^2\ln\frac{M^2}{\kappa^2}\right]^{-1}.
    \end{split}
    \end{equation}
    In our model, the field mixing phenomenon appears starting  from the one-loop level. Hence we have that $\tan{2\theta_{mix}}\sim g_{\phi}g_{\chi}$, $\sin{2\theta_{mix}}\sim g_{\phi}g_{\chi}$ and $\cos{2\theta_{mix}}\sim 1$. Therefore, sine and tangent of the $2\theta_{mix}$ are quantities of the second order in expansion series over the Yukawa couplings. Because of this, we in what follows  omit such quantities as $g_{\phi}^2\tan{2\theta_{mix}}$, $g_{\phi}g_{\chi}\sin{2\theta_{mix}}$ etc, which correspond to  next-to-leading corrections.

    Moreover,  the mixing angle roughly is  proportional to $(\Lambda - \mu)^{-1}$. So, if one of the bosons  is much heavier than another, i.e. when $\mu \ll \Lambda$ or $\mu \gg \Lambda$, then $\theta_{mix}$ becomes very small. In this case, two particle resonances are located far one from another. As a result, the properties of the dark boson $\chi$ do not affect the mass or width of the visible particle $\phi$, and the dark matter remains invisible. Alternatively, if we put $g_{\chi}\ll g_{\phi}$, then $\theta_{mix}$ may be very small and the dark matter signal will be indistinguishable from the  background of visible particle. Thus, the weak interplay of two sectors in the model corresponds to small values of the mixing angle between $\phi$ and $\chi$ states.

    \section{Width of  DM particles}

\begin{figure}[h]
    	\vskip1mm
    	\begin{center}
    		\includegraphics[scale=0.5]{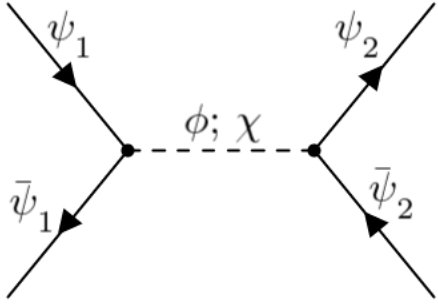}
    	\end{center}
    	\vskip-3mm
    	\caption{Feynman diagram of the investigated process. Bold points denote the vertexes calculated in one-loop aproximation. Dashed line corresponds to the two-point Green function of scalar doublet with the one-loop corrections.}
    	\label{fig:general-process}
    \end{figure}

\begin{figure}[h]
    	\vskip1mm
    	\begin{center}
    		\includegraphics[scale=1.0]{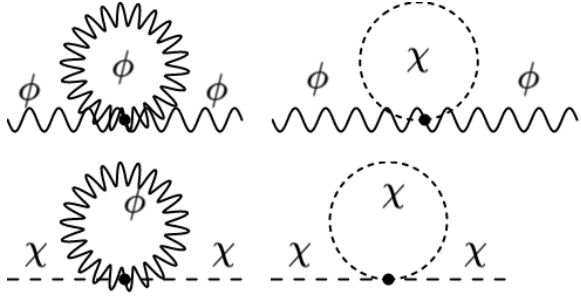}
    	\end{center}
    	\vskip-3mm
    	\caption{Loop corrections from the scalar fields self-interaction.}
    	\label{fig:scalar-loops}
    \end{figure}

    Let us consider the scattering process shown in Fig.\,\ref{fig:general-process}.
    \begin{figure}[h]
    	\vskip1mm
    	\begin{center}
    		\includegraphics[scale=0.5]{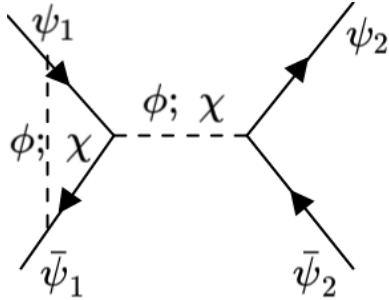}
    	\end{center}
    	\vskip-3mm
    	\caption{General view of the loop correction to the Yukawa interaction constants $g_{\phi}$ or $g_{\chi}$.  All lines and vertexes are taken in tree-approximation.}
    	\label{fig:yukawa-vertex-correction}
    \end{figure}
    In this reaction, a pair of visible fermions $\psi_1$ and $\bar{\psi}_1$ annihilates into other visible fermion pair $\psi_2$ and $\bar{\psi}_2$. As an intermediate state, we have both $\phi$ and $\chi$ bosons. Besides the diagrams in fig.\;\ref{fig:one-loop-mixing}, we take into account the contributions coming from the one-loop vertex corrections arising from the diagrams such as that in fig. \ref{fig:yukawa-vertex-correction}. The matrix element of the process in fig.\,\ref{fig:general-process} in improved Born approximation has the following expression:
    \begin{equation}
    \label{matrix-element}
    \begin{split}
    i\mathcal{M}^{(imp.\,Born)} =  & i \left(\begin{array}{c}
    g_{\phi}\left(1 + \delta\Gamma_{\phi}^{(fin)}(p_3;p_4)\right) \\
    g_{\chi}\left(1 + \delta\Gamma_{\chi}^{(fin)}(p_3;p_4)\right)
    \end{array}\right)^T \left(\begin{array}{cc}
    p^2 - \mu^2 - \Pi_{\phi\phi}(p^2) & - \Pi_{\phi\chi}(p^2) \\ - \Pi_{\phi\chi}(p^2) & p^2 - \Lambda^2 - \Pi_{\chi\chi}(p^2)
    \end{array}\right)^{-1} \\
    & \times\left(\begin{array}{c}
    g_{\phi}\left(1 + \delta\Gamma_{\phi}^{(fin)}(p_1;p_2)\right) \\
    g_{\chi}\left(1 + \delta\Gamma_{\chi}^{(fin)}(p_1;p_2)\right)
    \end{array}\right).
    \end{split}
    \end{equation}
    Here $\delta\Gamma_{\phi}^{(fin)}$ and $\delta\Gamma_{\chi}^{(fin)}$ denote the renormalized loop corrections depicted in the diagram \ref{fig:yukawa-vertex-correction}, which we calculate numerically\footnote{Hereafter, all numerical simulation results were obtained by the means of the LoopTools \cite{bibl:hahn} software.}. $\Pi_{\phi\phi}(p^2)$, $\Pi_{\phi\chi}(p^2)$ and $\Pi_{\chi\chi}(p^2)$ are the renormalized components of the scalar field polarization operators. They comprise  the contributions from the diagrams in fig.\,\ref{fig:one-loop-mixing} and those from the self-interaction of $\phi$ and $\chi$, shown in fig.\,\ref{fig:scalar-loops}. The scalar polarization operator components have been renormalized by using the following conditions:
    \begin{equation}
    \label{renorm-conditions}
    \begin{split}
    &\Re~\Pi_{\phi\phi}(p^2 = \mu^2) = 0,\quad\Re~\Pi_{\chi\chi}(p^2 = \Lambda^2) = 0,\quad\Re~\Pi_{\phi\chi}(p^2 = \kappa^2) = 0.
    \end{split}
    \end{equation}
    Here, diagonal terms $\Pi_{\phi\phi}(p^2)$ and $\Pi_{\chi\chi}(p^2)$ were renormalized on the mass shells of  corresponding bosons. For the non-diagonal component $\Pi_{\phi\chi}(p^2)$, an arbitrary   renormalization point $\kappa^2$ was used.
    It is worth noting that in the one-loop approximation we would also have to take into consideration the box diagrams of the process in fig.\,\ref{fig:general-process}. However, numerical simulations show that their contribution to the squared modulus of the total matrix element is less than $1\%$ of the $|\mathcal{M}^{(imp.\,Born)}|^2$. So, we neglect the box diagrams in the $\chi$ boson width analysis. We also found that loop corrections $|\delta\Gamma_{\phi}|\ll 1$ and $|\delta\Gamma_{\chi}|\ll 1$ for the considered area of the model parameter space, and the order of their absolute values do not exceed $10^{-2}$. Due to this, the matrix element \eqref{matrix-element} can be simplified to the following form:
    \begin{equation}
    \label{simplified-matrix-element}
    \begin{split}
    &i\mathcal{M}^{(imp.\,Born)} \approx i\left(\begin{array}{c}
    g_{\phi} \\
    g_{\chi}
    \end{array}\right)^T \left(\begin{array}{cc}
    p^2 - \mu^2 - \Pi_{\phi\phi}(p^2) & - \Pi_{\phi\chi}(p^2) \\ - \Pi_{\phi\chi}(p^2) & p^2 - \Lambda^2 - \Pi_{\chi\chi}(p^2)
    \end{array}\right)^{-1}\left(\begin{array}{c}
    g_{\phi} \\
    g_{\chi}
    \end{array}\right).
    \end{split}
    \end{equation}
    The matrix between the coupling constants vectors  is the two-point bosonic Green function in  momentum representation. Its diagonal elements correspond to the functions $\left<0\left|T\phi(x_1)\phi(x_2)\right|0\right>$ and $\left<0\left|T\chi(x_1)\chi(x_2)\right|0\right>$, while the off-diagonal element corresponds to $\left<0\left|T\phi(x_1)\chi(x_2)\right|0\right>$. We then express \eqref{simplified-matrix-element} in terms of the fields $\phi^{\prime}$ and $\chi^{\prime}$, which are rotated on the angle $\theta_{mix}$ with respect to the $\phi$ and $\chi$, and obtain the following diagonalized matrix element,
    \begin{equation}
    \label{diagonalized-matrix-element}
    \begin{split}
    &i\mathcal{M}^{(diag.)} = \frac{i\left(g_{\phi}\cos{\theta_{mix}} + g_{\chi}\sin{\theta_{mix}}\right)^2}{\left(p^2 - \mu^2 - \Pi_{\phi\phi}(p^2)\right)\left(1 - \frac{\Pi_{\chi\chi}(p^2)}{p^2 - \Lambda^2}\right)} + \frac{i\left(- g_{\phi}\sin{\theta_{mix}} + g_{\chi}\cos{\theta_{mix}}\right)^2}{\left(p^2 - \Lambda^2 - \Pi_{\chi\chi}(p^2)\right)\left(1 - \frac{\Pi_{\phi\phi}(p^2)}{p^2 - \mu^2}\right)} + O(g^4).
    \end{split}
    \end{equation}
    In this expression, all contributions having higher orders in  powers of the Yukawa couplings are denoted as $O(g^4)$. This term also depends on the external momentum. Hence, the rotation of the scalar fields on the angle $\theta_{mix}$ diagonalizes the matrix element \eqref{simplified-matrix-element}  in the second order in the Yukawa couplings, whereas the non-diagonal terms are proportional to the higher powers of the couplings.

    The matrix element \eqref{diagonalized-matrix-element} corresponds to the scattering process where interaction is mediated through the "physical" bosons $\phi^{\prime}$ and $\chi^{\prime}$, instead of the initial particles $\phi$ and $\chi$. That is, the resonances in the corresponding cross-section are formed out of the $\phi^{\prime}$ and $\chi^{\prime}$ for $p^2 = \mu^2$ and $p^2 = \Lambda^2$. The first and the second terms in \eqref{diagonalized-matrix-element} correspond to the $\phi^{\prime}$ and $\chi^{\prime}$ exchange, respectively. We denote these terms as $\mathcal{M}^{(\phi)}$ and $\mathcal{M}^{(\chi)}$:
    \begin{equation}
        \label{diagonalized-matrix-element-decomposition}
        \begin{split}
            & i\mathcal{M}^{(diag.)} \approx i\mathcal{M}^{(\phi)} + i\mathcal{M}^{(\chi)}, \\
            & i\mathcal{M}^{(\phi)} = \frac{i\left(g_{\phi}\cos{\theta_{mix}} + g_{\chi}\sin{\theta_{mix}}\right)^2}{\left(p^2 - \mu^2 - \Pi_{\phi\phi}(p^2)\right)\left(1 - \frac{\Pi_{\chi\chi}(p^2)}{p^2 - \Lambda^2}\right)}, \\
            & i\mathcal{M}^{(\chi)} = \frac{i\left(- g_{\phi}\sin{\theta_{mix}} + g_{\chi}\cos{\theta_{mix}}\right)^2}{\left(p^2 - \Lambda^2 - \Pi_{\chi\chi}(p^2)\right)\left(1 - \frac{\Pi_{\phi\phi}(p^2)}{p^2 - \mu^2}\right)}.
        \end{split}
    \end{equation}
    Thus, the width $\Gamma$ of the $\chi^{\prime}$ resonance is determined by the imaginary part of the polarization operator component $\Pi_{\chi\chi}(p^2)$ and reads
    \begin{equation}
    \label{chi-width}
    \begin{split}
    &\Gamma \approx \frac{\Im\Pi_{\chi\chi}(p^2 = \Lambda^2)}{\Lambda} = \frac{G_{\chi}^2\Lambda}{8\pi}\left(1 - \frac{4M^2}{\Lambda^2}\right)^{\frac{3}{2}} + \frac{g_{\chi}^2\Lambda}{8\pi}\left[\left(1 - \frac{4m_1^2}{\Lambda^2}\right)^{\frac{3}{2}} + \left(1 - \frac{4m_2^2}{\Lambda^2}\right)^{\frac{3}{2}}\right] + O(g^4).
    \end{split}
    \end{equation}

\begin{figure}
    	\begin{center}
    		\begin{tabular}{cc}
    			\includegraphics[scale=0.5]{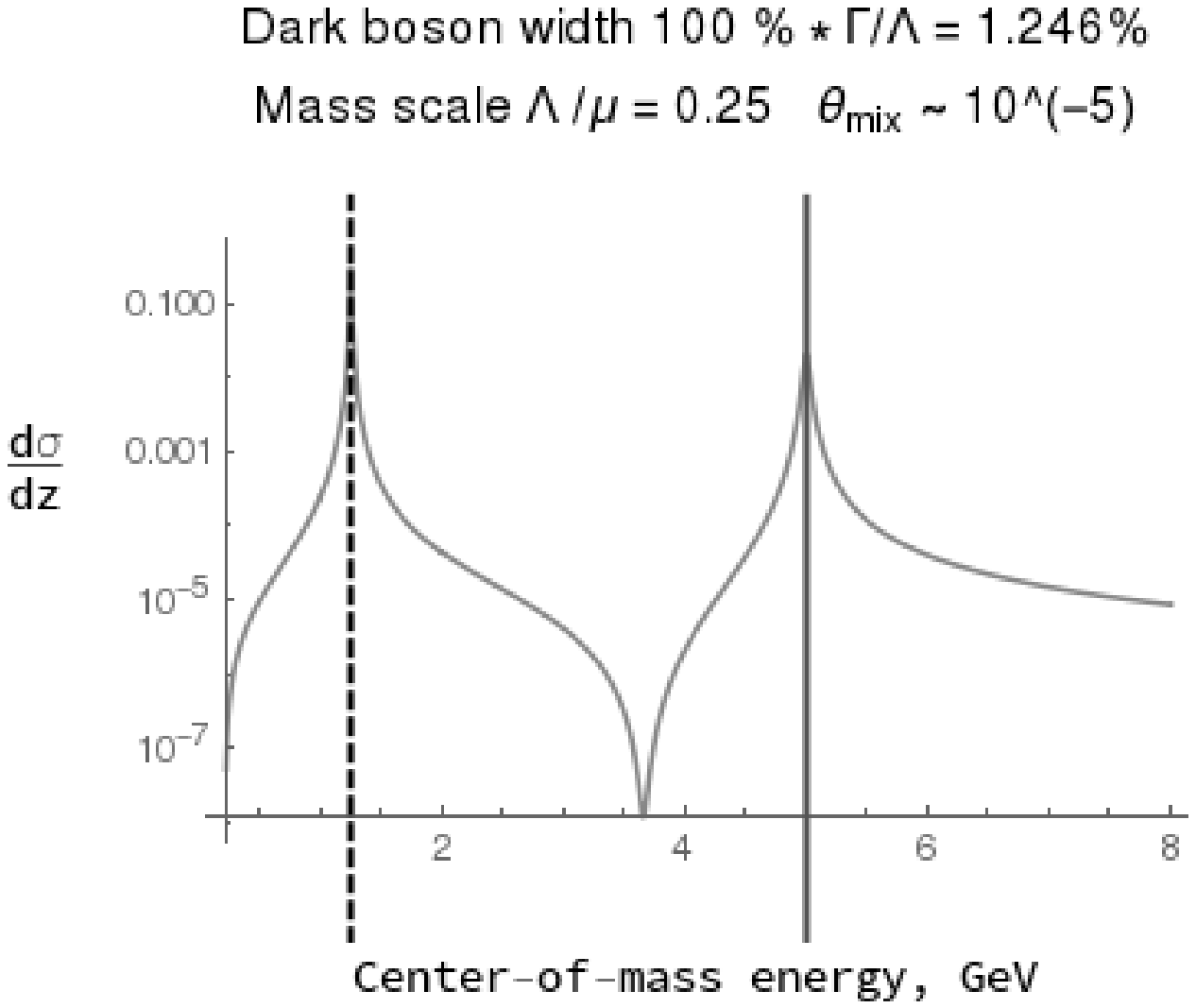} &
    			\includegraphics[scale=0.5]{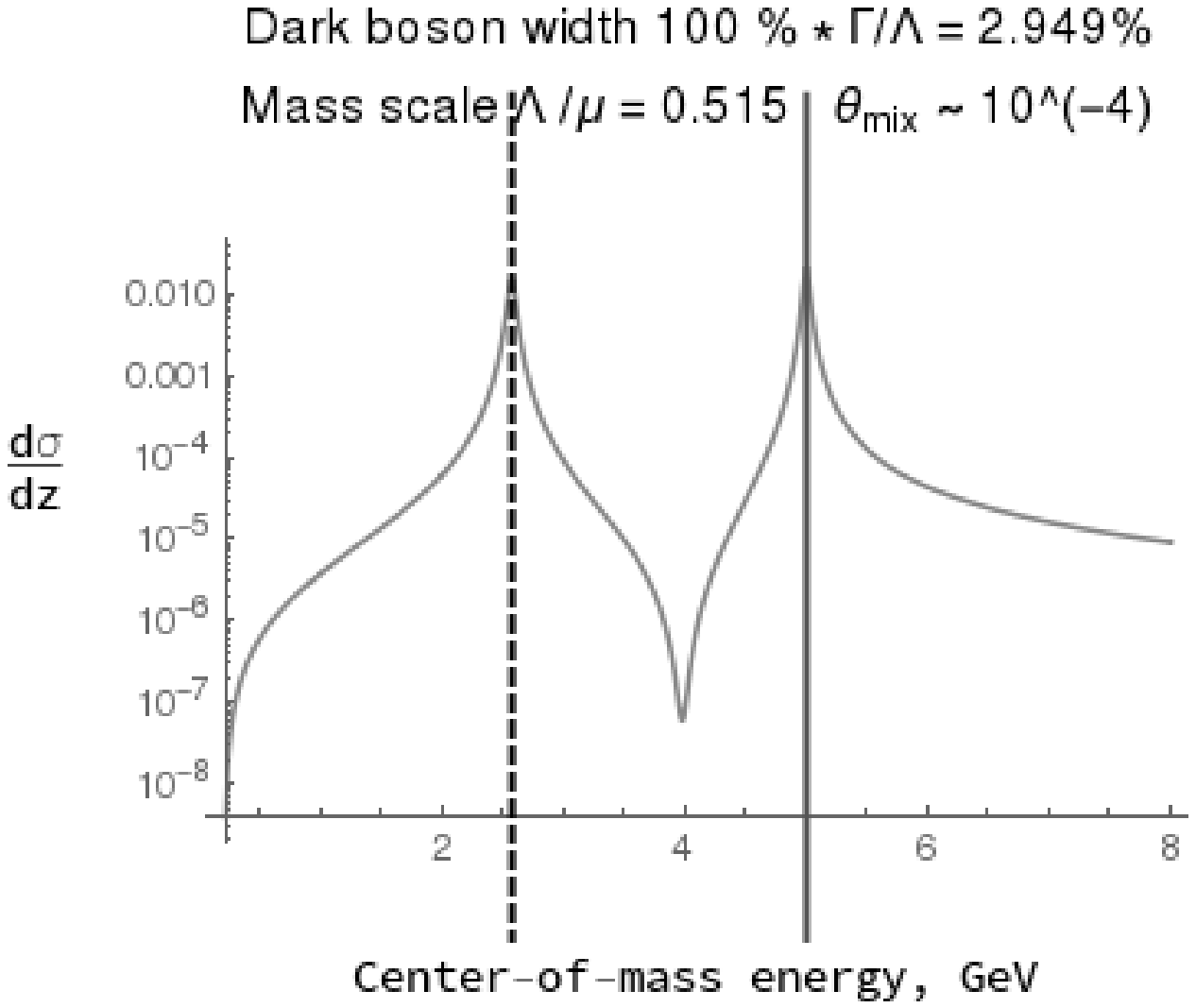} \\
    			\includegraphics[scale=0.5]{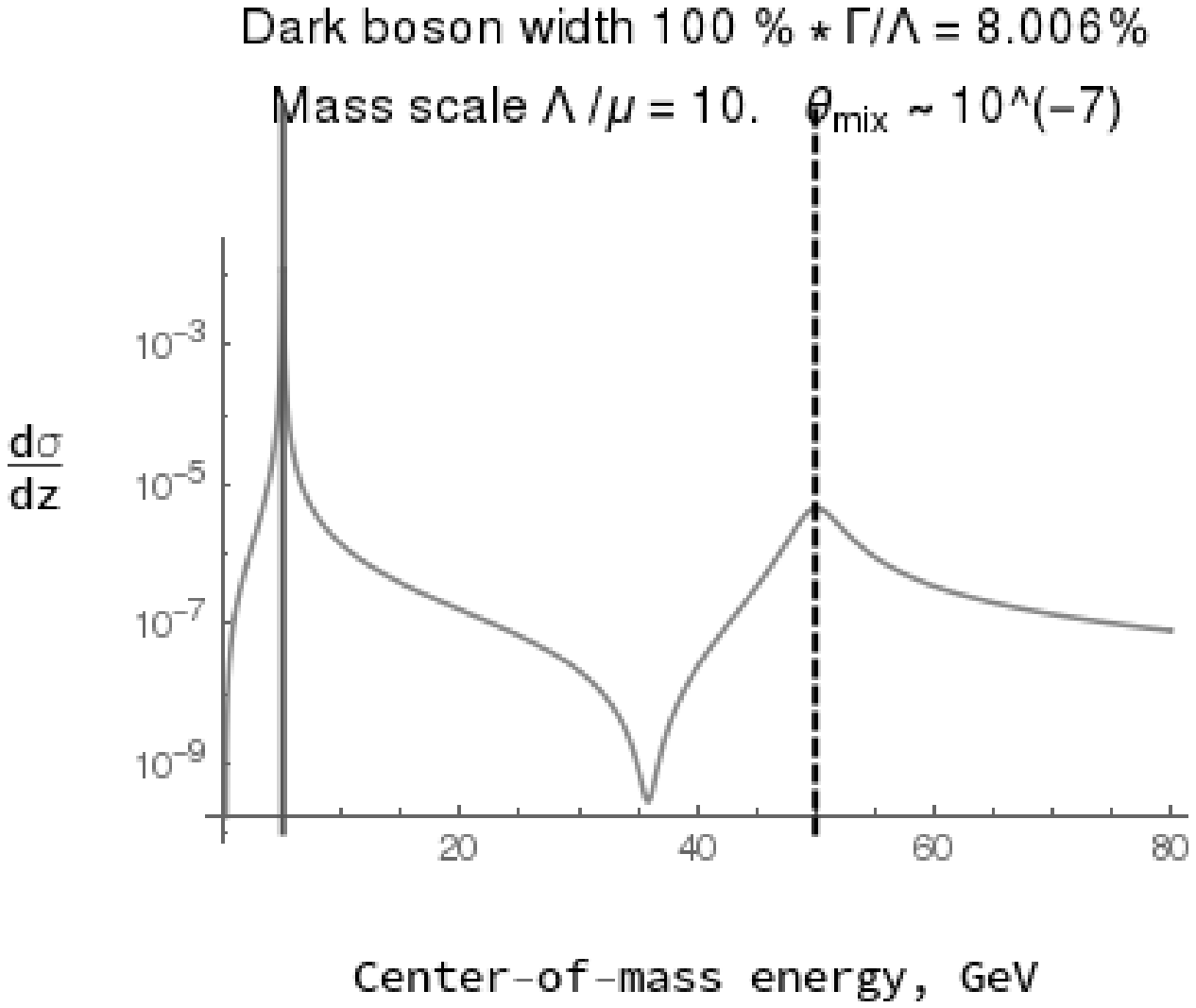}
    		\end{tabular}
    	\end{center}
    	\caption{Invariant mass spectra for the reaction products in fig.\;\ref{fig:general-process} for various $\Lambda$. $\Lambda$ values are given as fractions of the fixed $\mu$ mass. Peak positions are denoted as solid line -- visible $\phi$, dashed line -- dark $\chi$}
    	\label{fig:width-dep:a}
    \end{figure}

\begin{figure}
    	\begin{center}
    		\begin{tabular}{cc}
    			\includegraphics[scale=0.5]{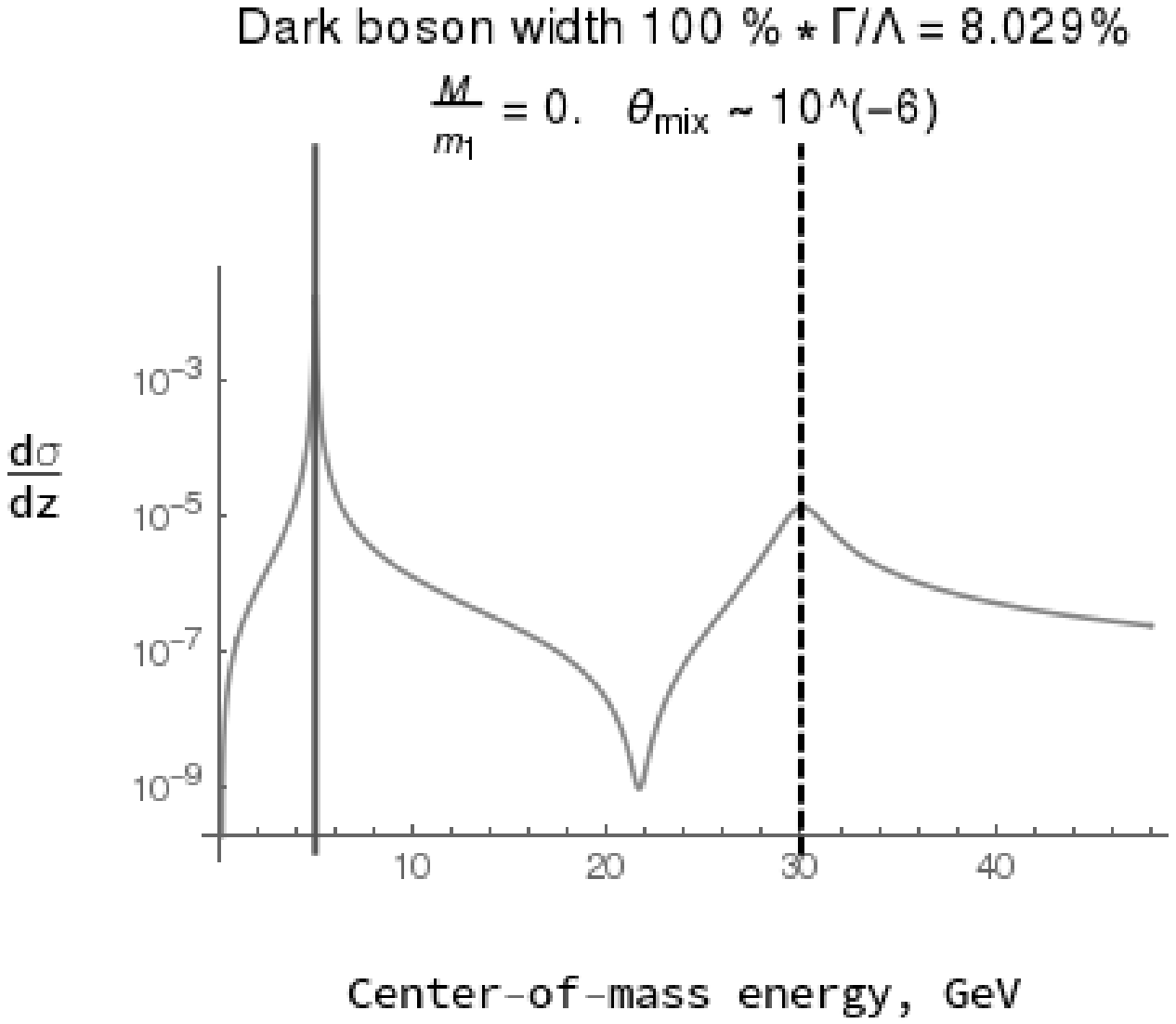} &
    			\includegraphics[scale=0.5]{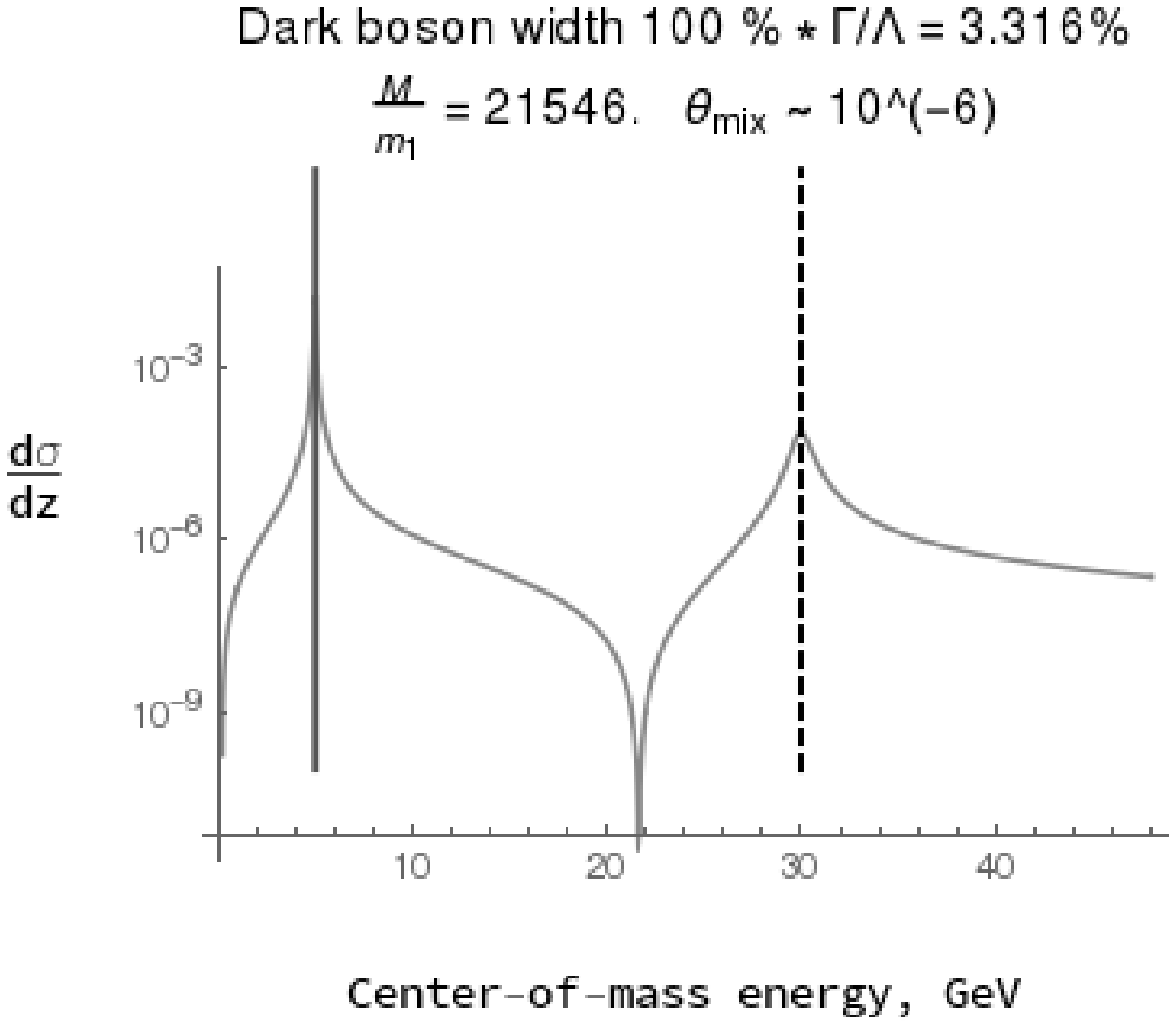}
    		\end{tabular}
    	\end{center}
    	\caption{Invariant mass spectra for the reaction products in fig.\;\ref{fig:general-process} for various $M$. Here, $\Lambda>\mu$ and $G_{\chi} \approx 3g_{\phi}$. Peak positions are denoted as solid line -- visible $\phi$, dashed line -- dark $\chi$}
    	\label{fig:width-dep:b}
    \end{figure}

    As free parameters  we have the dark boson mass $\Lambda$, couplings $g_{\chi}$, $G_{\chi}$, mixing angle $\theta_{mix}$ and dark fermion mass $M$. These quantities are the most interesting, since they contain information about properties of dark sector, its interaction with visible matter. And these values affect the width of the dark boson. In the model, we have such ranges of these parameters, where the resonance of $\chi$ becomes too wide to be described by the NWA. The invariant mass spectra for some  values of the parameters and corresponding dark peak widths are depicted in fig.\ref{fig:width-dep:a}, \ref{fig:width-dep:b}. In these figures, for convenience of readers we use a logarithmic scale along the axis $OY$.  The  width of visible particle is always small and lies in the range $\sim 1-2\%$. Vertical axis in the figures denotes differential cross-section $\frac{d\sigma}{dz}$ value, where $z=2\pi\cos\theta$ and $\theta$ is an angle between directions of initial $\psi_1$ and final $\psi_2$. The differential cross-section is given in $GeV^{-2}$ in all figures across this paper.


    We make conclusion about the validity of the NWA, by applying it to the DM resonances having different parameter values\footnote{Detailed data behind such estimate is given in Appendix B, in tables \ref{table:lambda-var} \ref{table:darkm-var}, \ref{table:gchi-var} and \ref{table:dark-gchi-var}}. Generally, the NWA-estimated contribution $\sigma_d^{(NWA)}$ of the dark resonance into the total cross-section deviates significantly from its value $\sigma_d$ numerically integrated over $p^2$ when $\chi$ width rises beyond $6-7\%$. This value has to be considered as the  upper width limit for the NWA applicability in our model. This  bound  is in addition to the $3\%$-limit, introduced by  the experimental data treatment techniques.

\begin{figure}[h!]
    	\vskip1mm
    	\begin{center}
    		\begin{tabular}{cc}
    			\includegraphics[scale=0.5]{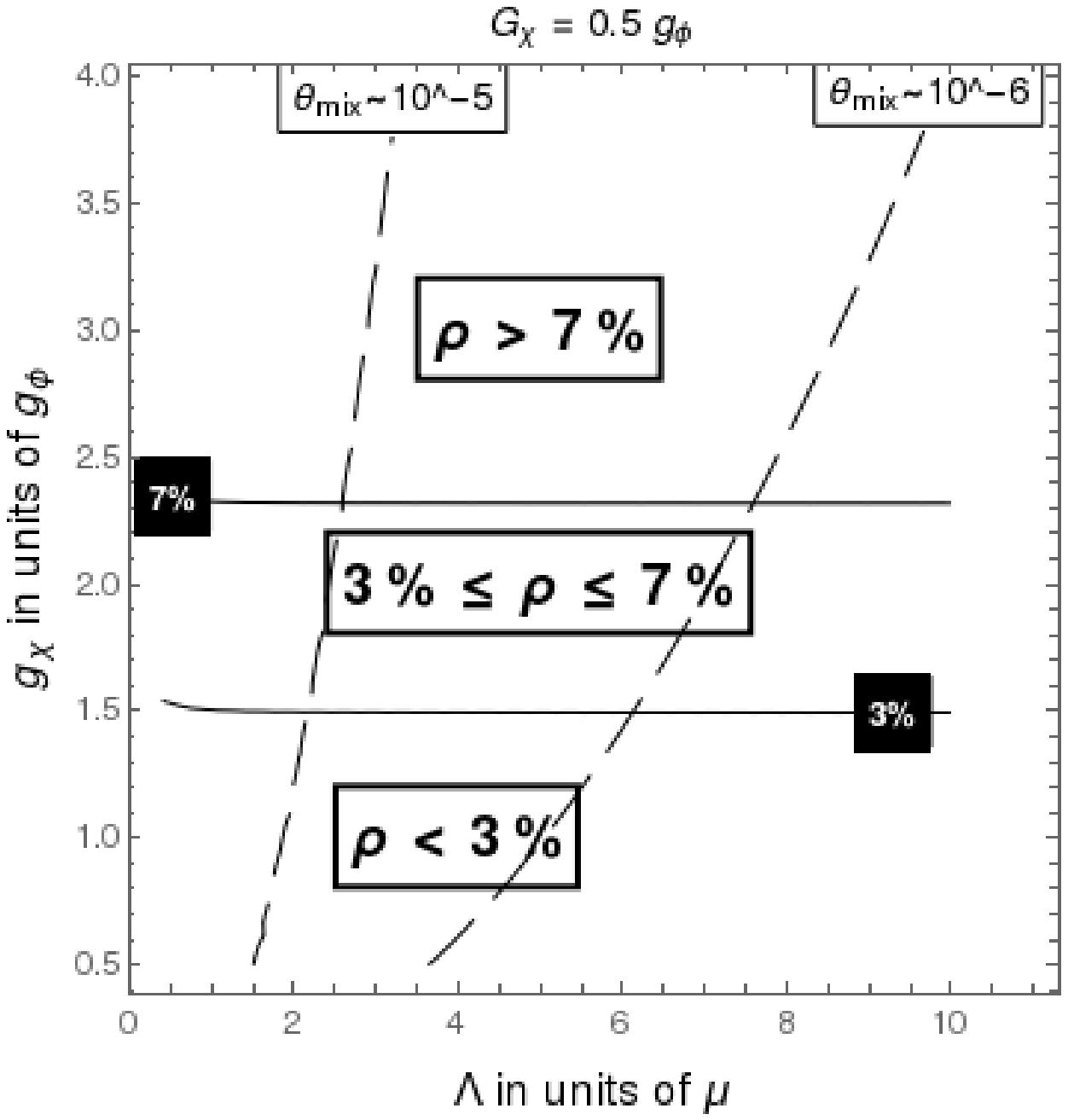} &
    			\includegraphics[scale=0.5]{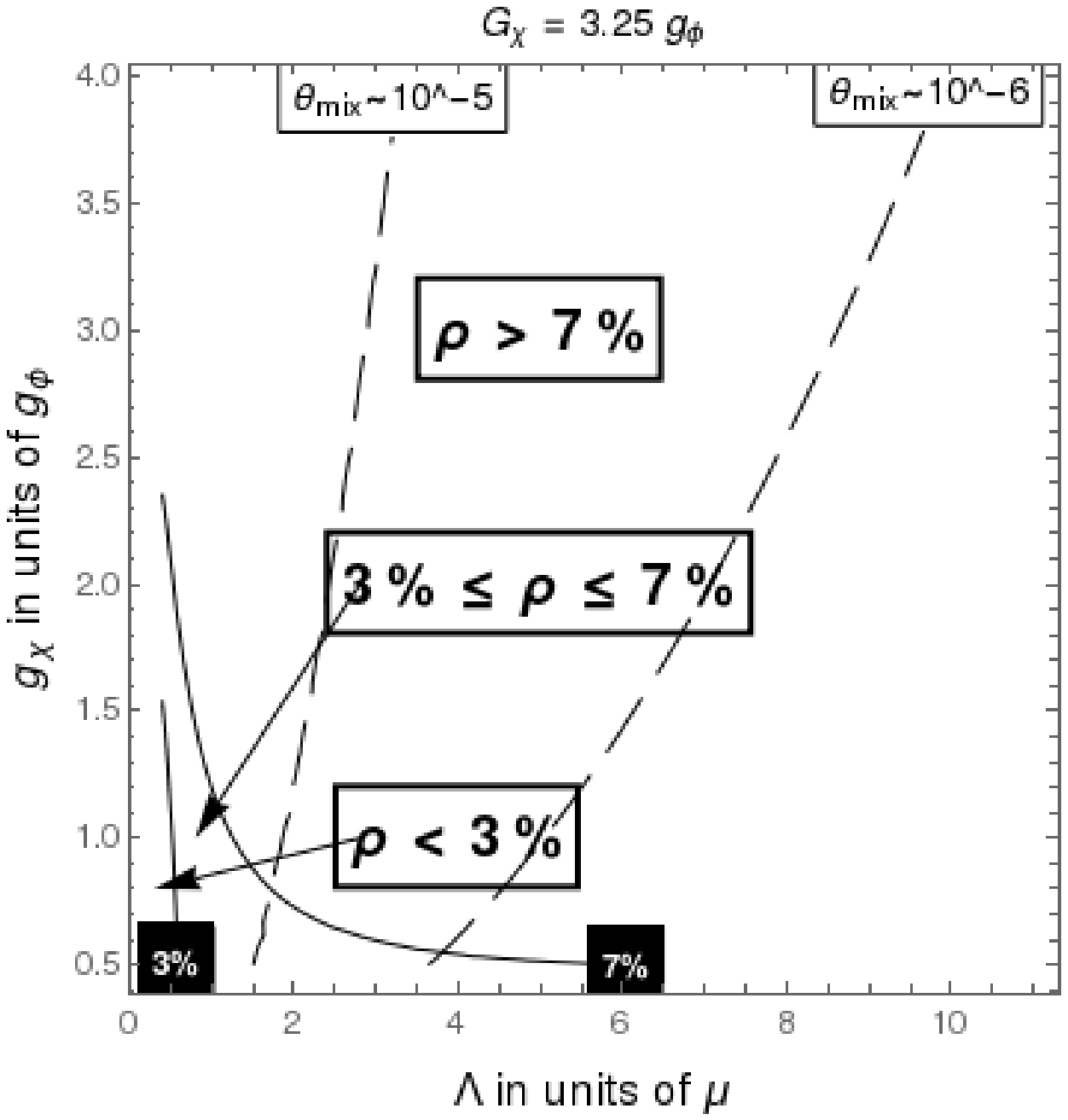}
    		\end{tabular}
    	\end{center}
    	\vskip-3mm
    	\caption{Countour map of dark boson widths and $\theta_{mix}$ magnitude levels in the  parametric space $\Lambda - g_{\chi}$, with fixed $G_{\chi}$. Contours corresponding to the different $\rho$ are subscribed with black labels, while $\theta_{mix}$ magnitude levels are subscribed with white labels.}
    	\label{fig:fixed-dark-gchi}
    \end{figure}

\begin{figure}[h!]
    	\vskip1mm
    	\begin{center}
    		\begin{tabular}{cc}
    			\includegraphics[scale=0.5]{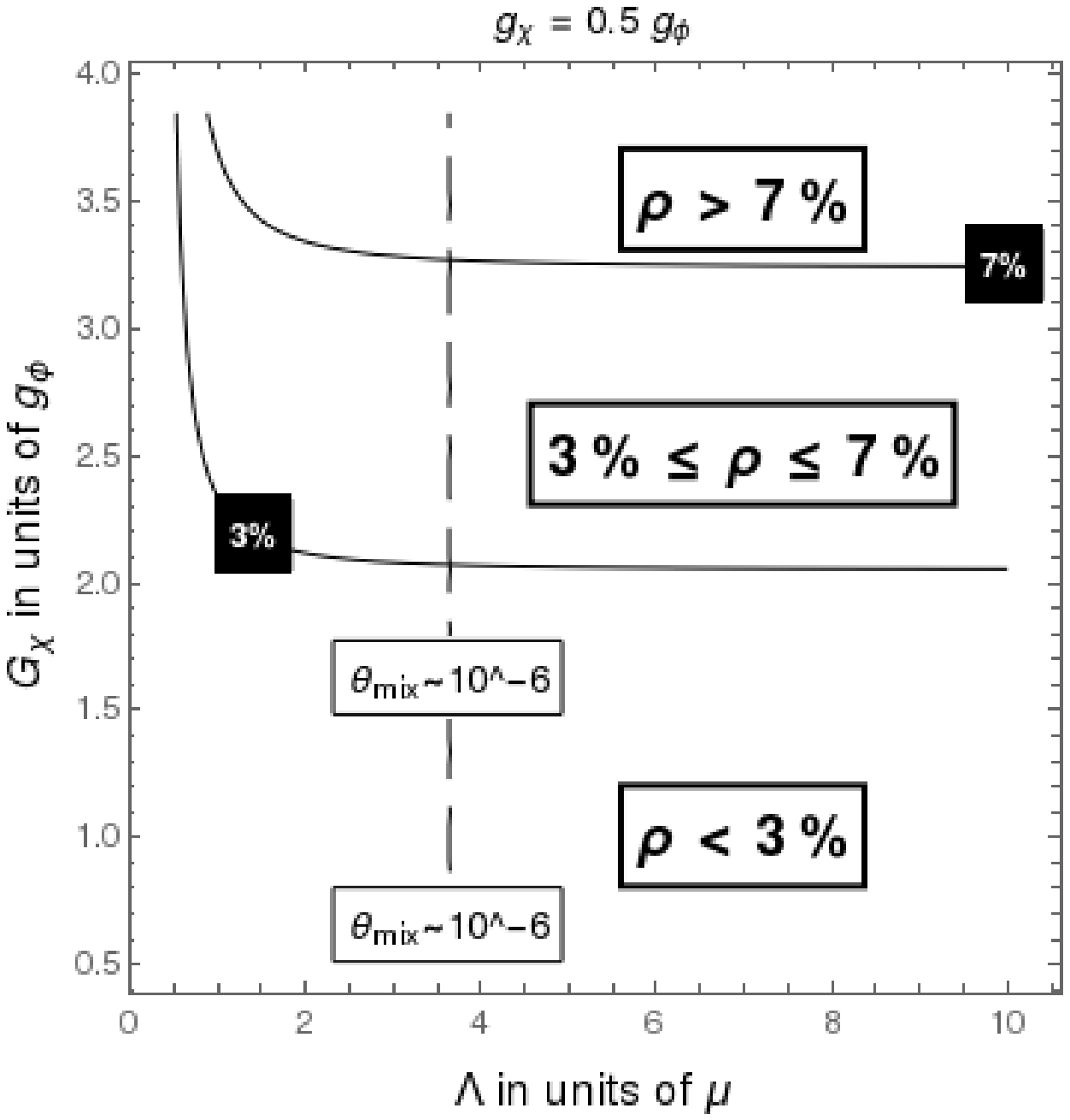} &
    			\includegraphics[scale=0.5]{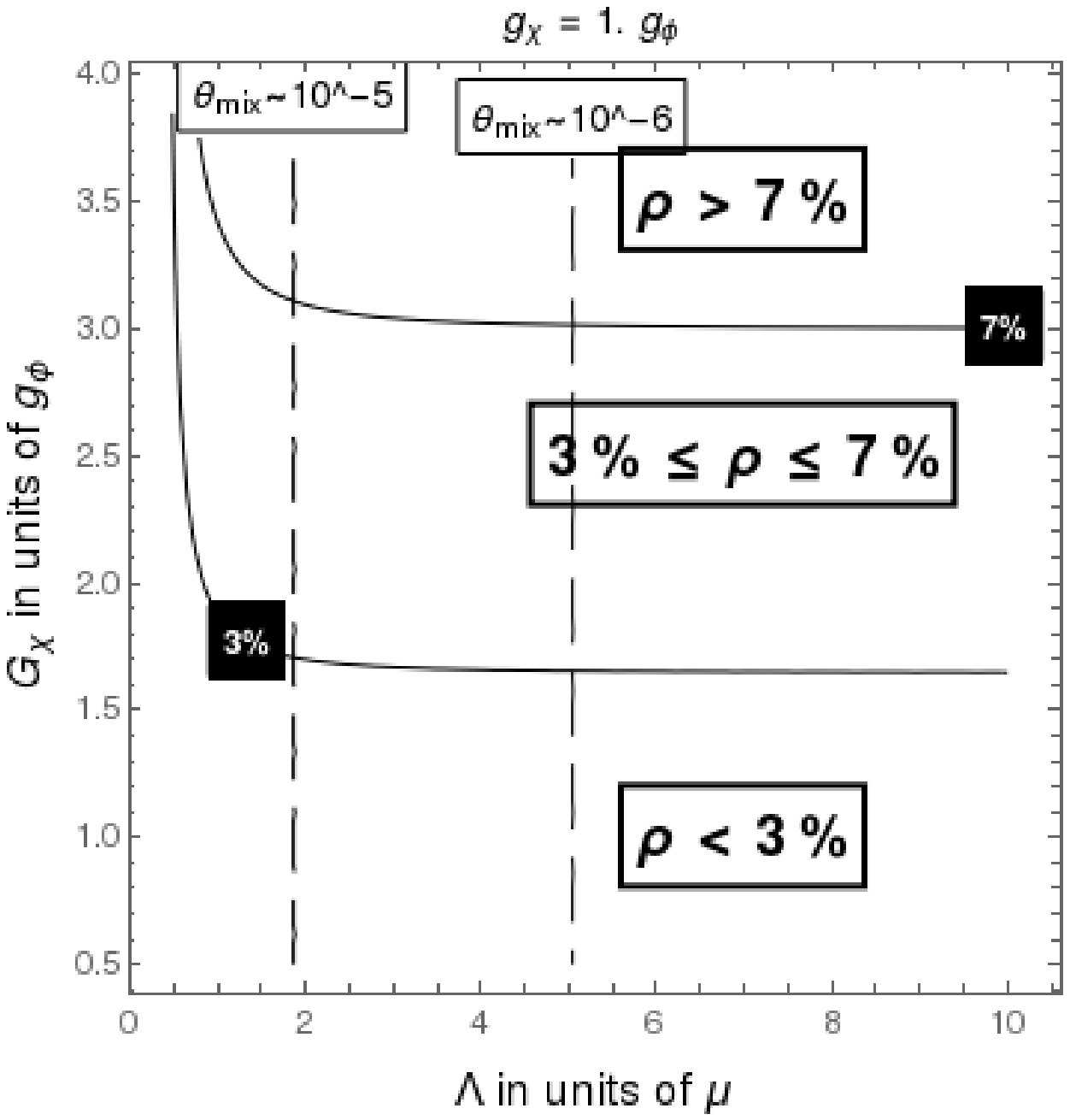} \\
    			\includegraphics[scale=0.5]{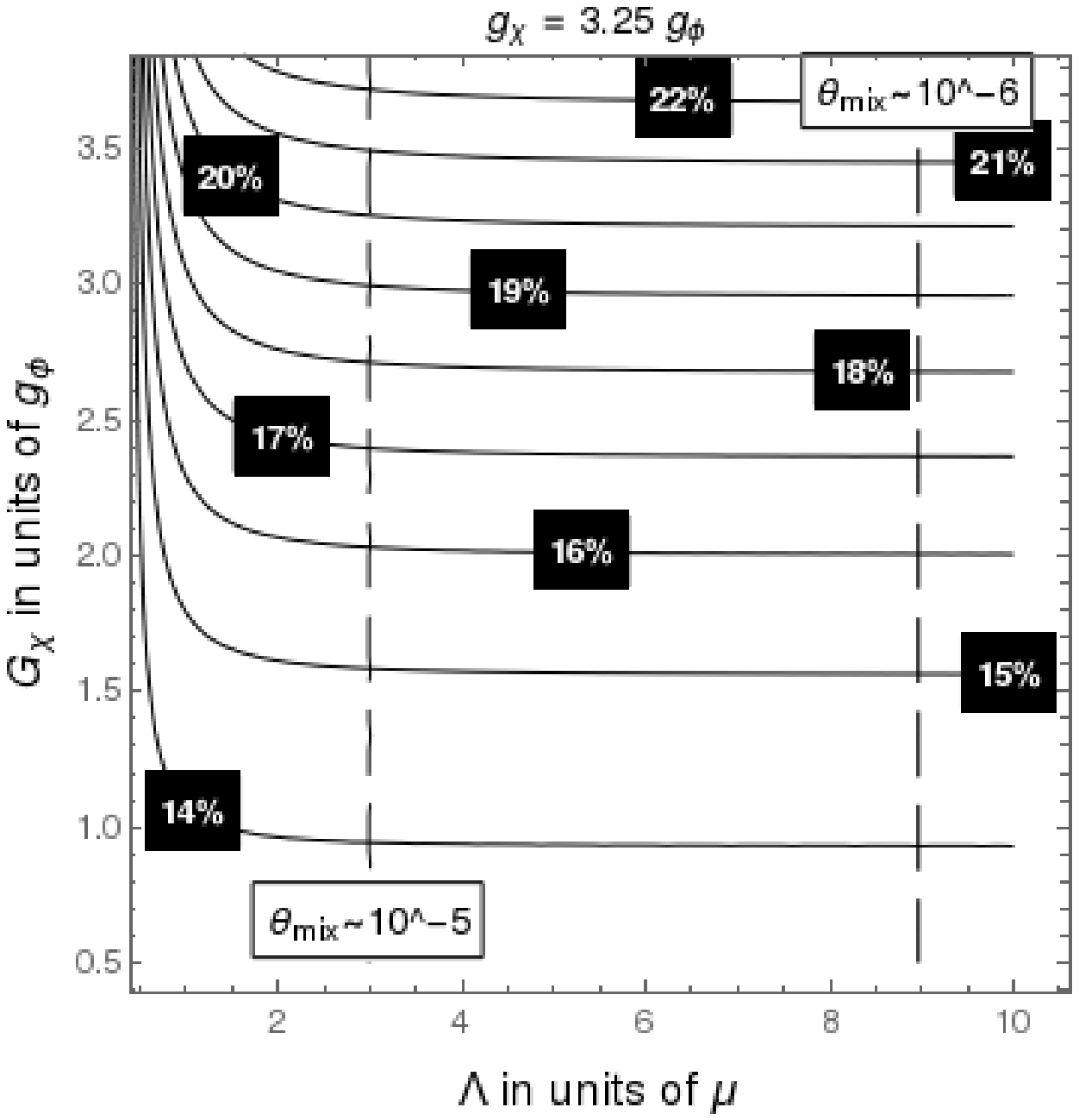}
    		\end{tabular}
    	\end{center}
    	\vskip-3mm
    	\caption{Contour  map of the dark boson widths and the $\theta_{mix}$ magnitude levels in the parametric space $\Lambda - G_{\chi}$, with fixed $g_{\chi}$. Contours corresponding to the different $\rho$ are subscribed with black labels, while $\theta_{mix}$ magnitude levels are subscribed with white labels.}
    	\label{fig:fixed-gchi}
    \end{figure}

    The results on the dark boson width calculation are collected in figs. \ref{fig:fixed-dark-gchi} and \ref{fig:fixed-gchi}. Widths presented there were calculated analytically, according to  formula \eqref{chi-width}\footnote{The validity of the analytical approximation \eqref{chi-width} was assessed via the numerical calculation of the dark resonance width, from the expression for the cross-section of the process in fig. \ref{fig:general-process}. The Yukawa vertex corrections (fig. \ref{fig:yukawa-vertex-correction}) were taken into account in this calculation. It was found that widths values from \eqref{chi-width} differ from the numerical calculations results on less than $2\%$ of the mass $\Lambda$. Hence, formula \eqref{chi-width} provides approximation of the $\chi$ width, which is good enough in the considered model parameters range.}. They are presented as contour maps of the $\chi$ widths with two specific contours corresponding to $3\%$ and $7\%$, shown as black solid lines. There are areas which contain the peaks having ratio $\rho \equiv \Gamma/\Lambda$  less than $3\%$. So, they are potentially visible in experiment. Areas where $\rho > 7\%$ contain wide states and show the widths beyond the applicability of the NWA. The last areas in the graphs contain widths which can still be captured through the NWA, although they are wider than $3\%$. Also, the values of mixing angle are present there. In all these figures the parameter $M$ is taken to be the same ($M\approx 2\cdot 10^3 m_1$).

    As we can  see from the second graph in fig.\;\ref{fig:fixed-dark-gchi}, if new boson interacts  sufficiently strongly  with other particles inside the  dark sector,  the formation  of new narrow resonance is rather exclusive than typical. Namely, this  is so when new particle is lighter than known one. Such boson is easily detectable, so this case could be rejected. For $G_{\chi} \gg g_{\phi}$ we find that if new particle is heavier than the visible one,  its peak has to be wide.  Contrary to this, if $\chi$ and $\Psi$ interact weakly, the new peak is narrow one in a wide range of its mass and for coupling $g_{\chi}\approx g_{\phi}$. In the latter case, if $\Lambda\leq \mu$, the NWA is applicable to the dark peak for almost whole range of $G_{\chi}$ variation (fig.\;\ref{fig:fixed-gchi}). But if the coupling $g_{\chi}\gg g_{\phi}$ the detection of dark resonance by the means of NWA is impossible. As  we can  see from the last graph in fig.\;\ref{fig:fixed-gchi}, the narrow peaks do not exist in that case. Thus, to  keep the  wide dark resonance, there has to be $\Lambda>\mu$ and either $g_{\chi} \gg g_{\phi}$ or $G_{\chi} \gg g_{\phi}$.

    According to results of  modern experiments, new hypothetical bosons beyond the Standard model do not change properties of known  resonances \cite{bibl:gulov}. It is so if  the masses of two resonances are far enough one from another, and they do not overlap. This condition is satisfied, in particular, when the mixing angle $\theta_{mix}$ is small enough. For our model, we find that the necessary condition has to be $\theta_{mix}\leq 10^{-5}$. The levels of the mixing angle magnitude are also shown in figs. \ref{fig:fixed-dark-gchi} and \ref{fig:fixed-gchi} as black dashed lines. As it is depicted in the first plot of fig.\;\ref{fig:fixed-gchi}, the region $\theta_{mix}>10^{-5}$ corresponds to $\Lambda=O(\mu)$. If it is so, the effects of new physics could be detected either in direct on-resonance searching for narrow dark peak, or through its influence on the properties of visible particles, if it is wide.

    To summarise, we have investigated the influence of the dark fermion mass $M$ on the $\chi$-particle width. It is worth noting that the existence of additional fermions beyond the visible sector is not obligatory. This is  because only the properties of the dark boson particle are under consideration. Nevertheless, the condition of the wide $\chi$ resonance leads to certain restrictions on the $\Psi$-field mass. As we  can see from the plots in fig.\;\ref{fig:width-dep:b}, when mass $M$ is increasing the DM peak is  narrowing. Hence, to keep the width of the $\chi$ bigger than $3\%$, we have to anticipate that the dark fermion  mass values are  to be  in the scale of visible ones. The estimate of the $M$ upper bound is,  $M\leq (10^3-10^4)m_1$. Remind also, we have considered the light enough visible fermions contributing to the boson widths, so that $2m_{1;2}\ll \Lambda$.

     \section{Dark boson contribution to the cross-section}

    As it was shown in the previous section, when mass $\Lambda \leq O(\mu)$ dark boson could be detected either as a narrow resonance or via its influence on the properties of  visible particles. Since no dark matter particles with such masses were found, we assume that $\Lambda \gg \mu$. Taking this constraint into consideration, we investigate the dark boson contribution to the process cross-section. In the center-of-mass frame, this reads
    \begin{equation}
    \label{cross-section}
    \begin{split}
    & \sigma\Bigr|_{CM} = \Phi(p^2)\left|\mathcal{M}^{(simpl.\,diag.)}\right|^2 = \sigma^{(\phi)} + \sigma^{(interf.)} + \sigma^{(\chi)},\quad \Phi(p^2) = \frac{1}{8\pi}\frac{p^2 - 4m_2^2}{p^2}\sqrt{(p^2 - 4m_1^2)(p^2 - 4m_2^2)}, \\
    & \sigma^{(\phi)} = \Phi(p^2)\left|\mathcal{M}^{(\phi)}\right|^2,\quad\sigma^{(interf.)} = 2\Phi(p^2)\Re\left\{\mathcal{M}^{(\phi)}\mathcal{M}^{(\chi)\dagger}\right\},\quad\sigma^{(\chi)} = \Phi(p^2)\left|\mathcal{M}^{(\chi)}\right|^2.
    \end{split}
    \end{equation}
    Here $\sigma\Bigr|_{CM}$ consists of the resonant terms $\sigma^{(\phi)}$ and $\sigma^{(\chi)}$, and the interference term $\sigma^{(interf.)}$. These terms are as follows,
    \begin{equation}
    \label{cross-section-decomposition}
    \begin{split}
    & R(p^2) = \left[(p^2 - \mu^2)^2 + \mu^2\Gamma^{(vis.)2}\right]^{-1}\left[(p^2 - \Lambda^2)^2 + \Lambda^2\Gamma^{2}\right]^{-1}, \\
    & \sigma^{(\phi)} = \Phi(p^2)R(p^2)\left(g_{\phi}\cos\theta_{mix} + g_{\chi}\sin\theta_{mix}\right)^4(p^2 - \Lambda^2)^2, \\
    & \sigma^{(\chi)} = \Phi(p^2)R(p^2)\left(-g_{\phi}\sin\theta_{mix} + g_{\chi}\cos\theta_{mix}\right)^4(p^2 - \mu^2)^2, \\
    & \sigma^{(interf.)} = \frac{1}{2}\Phi(p^2)R(p^2)\left[2g_{\phi}g_{\chi}\cos{2\theta_{mix}} - (g_{\phi}^2 - g_{\chi}^2)\sin{2\theta_{mix}}\right]^2(p^2 - \mu^2)(p^2 - \Lambda^2).
    \end{split}
    \end{equation}
     $\Gamma^{(vis.)}$ stands for the width of $\phi$ boson.  Note that \eqref{diagonalized-matrix-element} approximates the matrix element \eqref{matrix-element} well when $p^2 \ll \Lambda^2$, $\Lambda^2 \gg \mu^2$. In this case the non-diagonal component $\Pi_{\phi\chi}(p^2)$ is small, and the  factorization of the matrix element  \eqref{diagonalized-matrix-element} is valid. For higher energies, the real parts of the polarization operator components rise, so they can not be neglected when $p^2 = O(\Lambda^2)$.

    For the model parameter values and the energies studied here the mean relative error between $\sigma\Bigr|_{CM}$ in \eqref{cross-section} and the cross-section in the improved Born approximation is less than $6\%$. In this estimation, the cross-section in the improved Born approximation has been calculated with the matrix element \eqref{matrix-element}.
   Now we turn to the interference contribution and
consider the energy range where $2\max\{m_1;m_2;M\} < \sqrt{p^2} \ll \Lambda$, and find such values of the  couplings for which the interference term is more significant compared the $\sigma^{(\chi)}$. As can be seen from \eqref{cross-section-decomposition}, in the region $\mu^2 < p^2 < \Lambda^2$ $\sigma^{(interf.)} < 0$ and $\sigma^{(\chi)} > 0$, so there might be such points where the interference term cancels the resonant term. In such cases, signal of the $\chi$ boson becomes very weak, and this particle can be missed in the experimental search. We identify the points where it is so.

\begin{figure}[h!]
    	\begin{center}
    		\begin{tabular}{cc}
    			\includegraphics[scale=0.48]{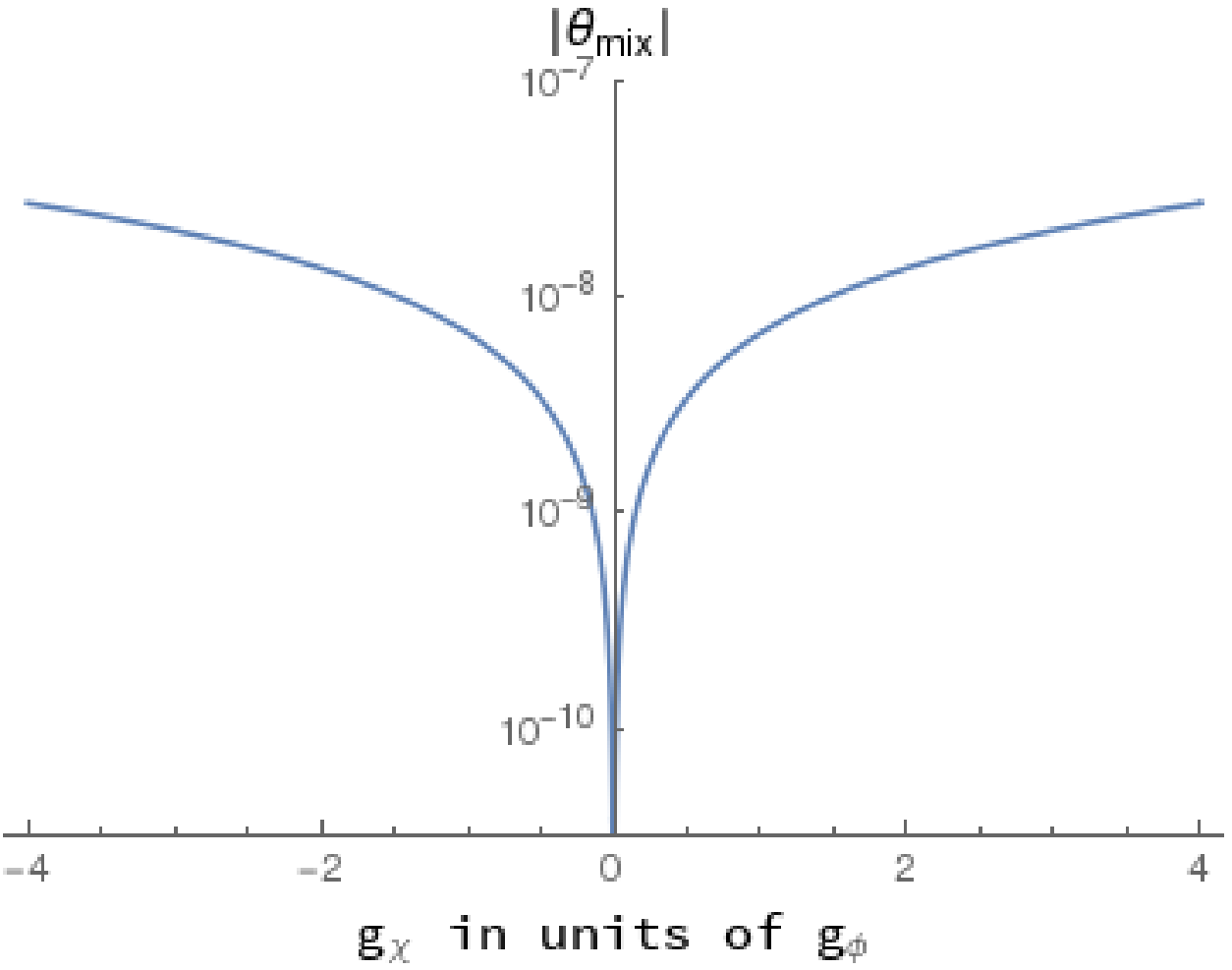} &
    			\includegraphics[scale=0.48]{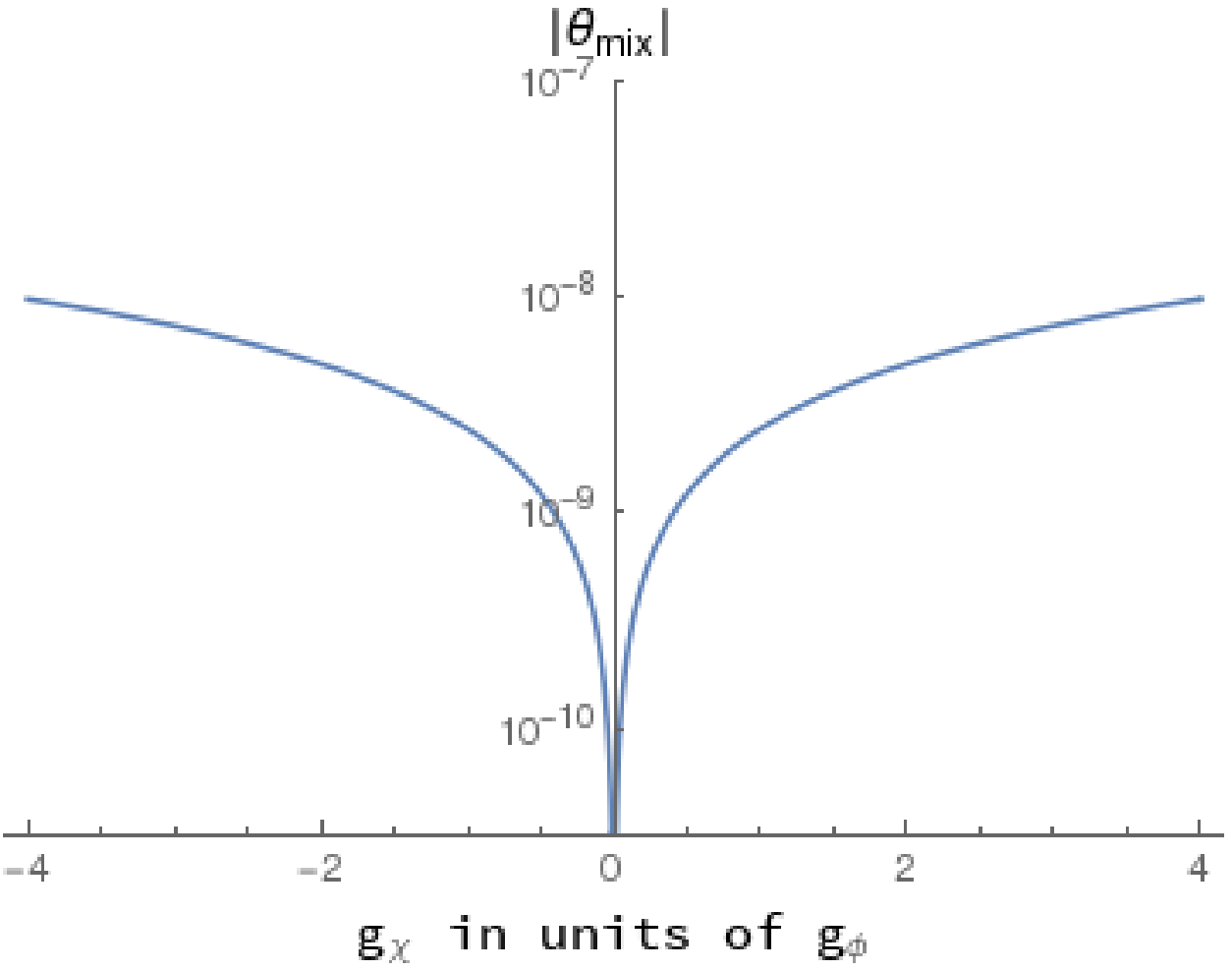} \\
    			\includegraphics[scale=0.48]{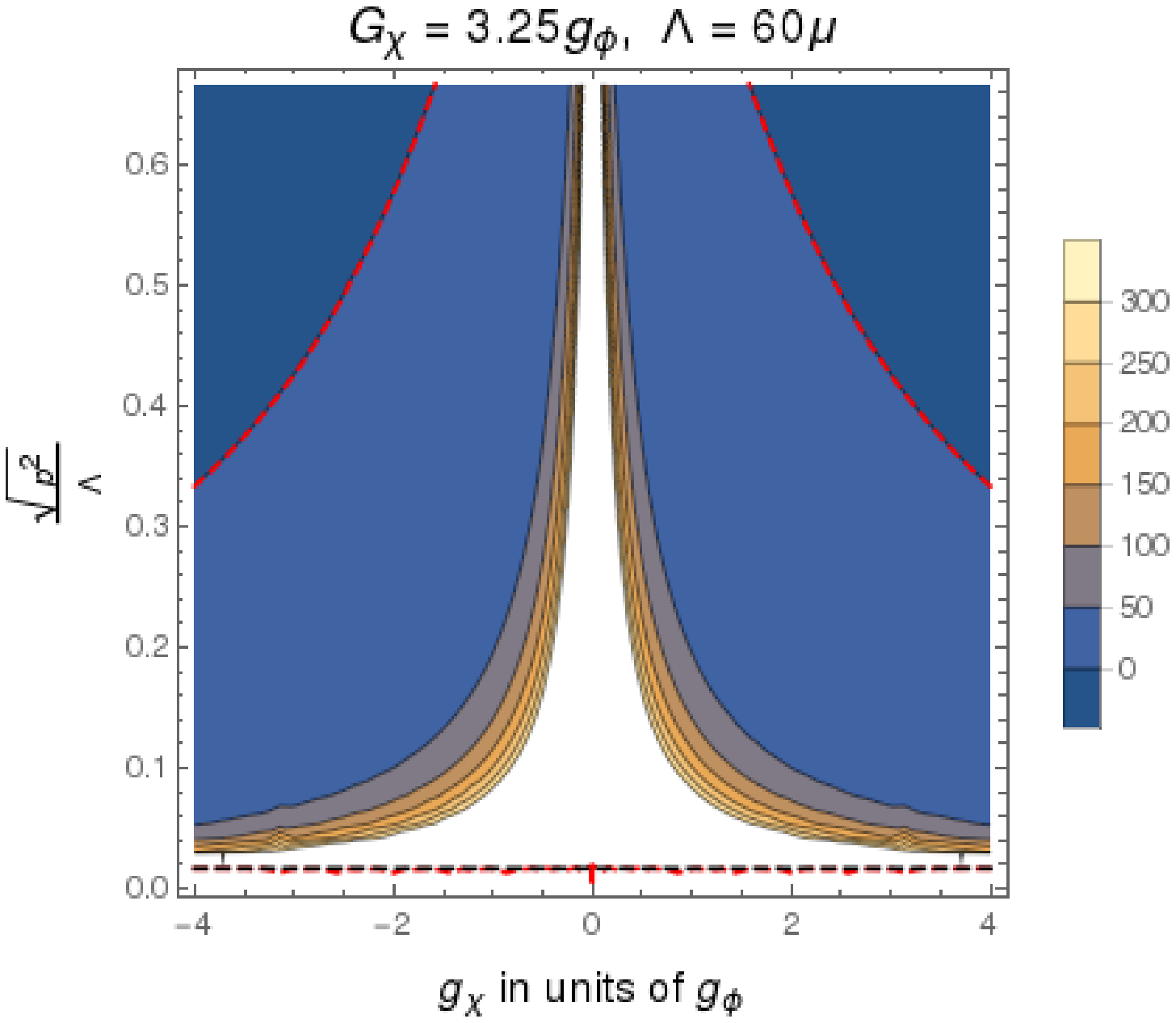} &
    			\includegraphics[scale=0.48]{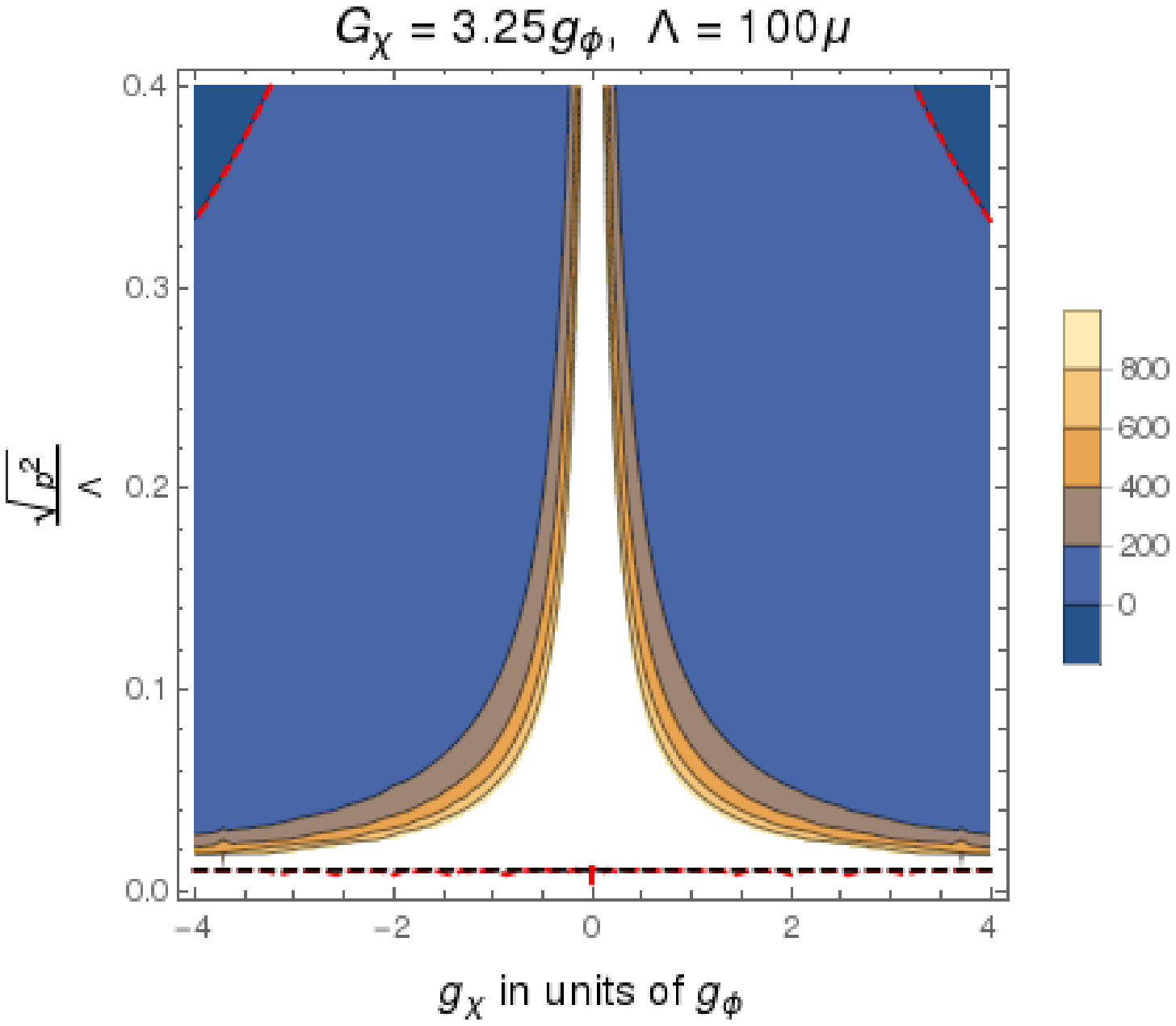} \\
    			\includegraphics[scale=0.48]{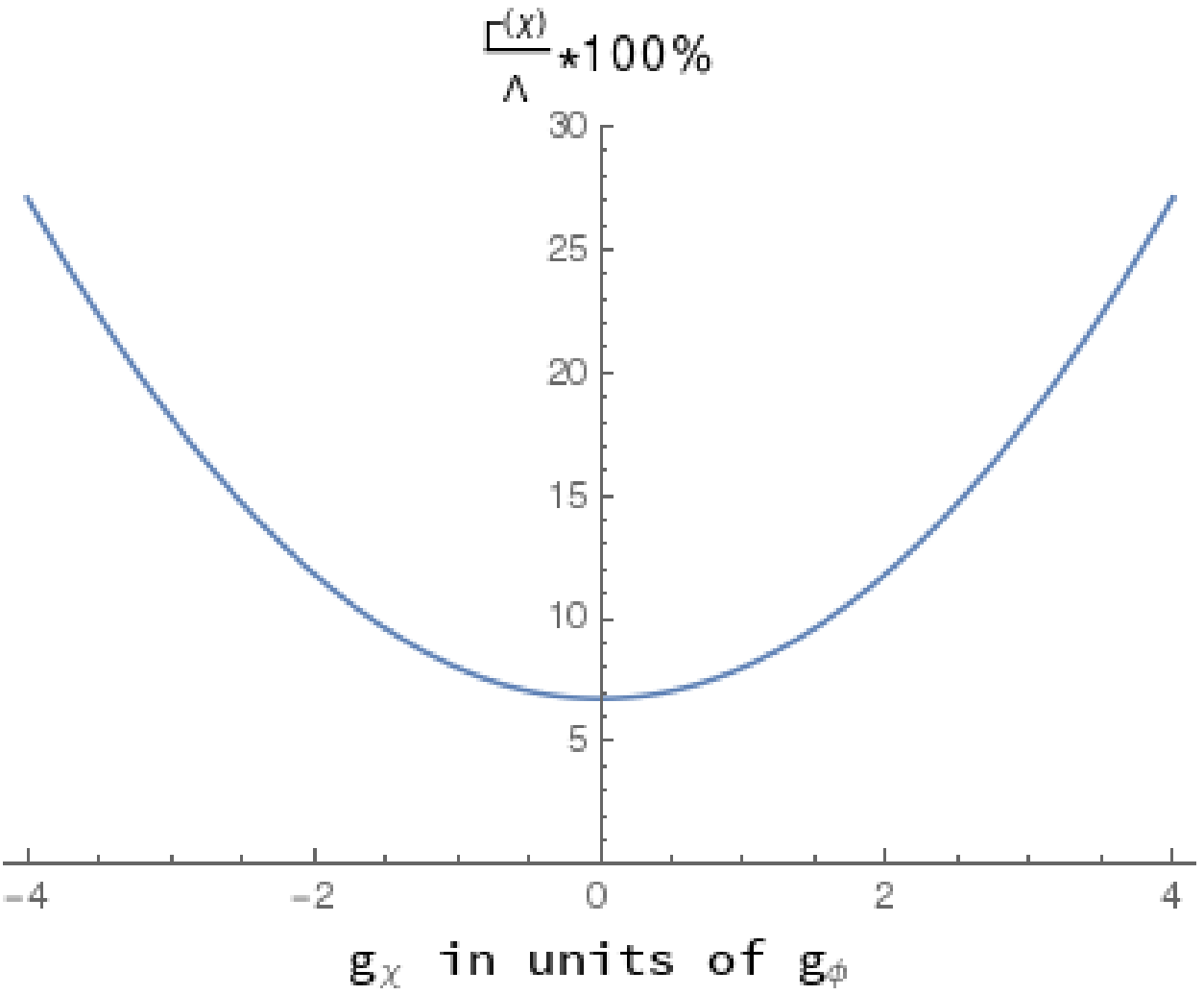} &
    			\includegraphics[scale=0.48]{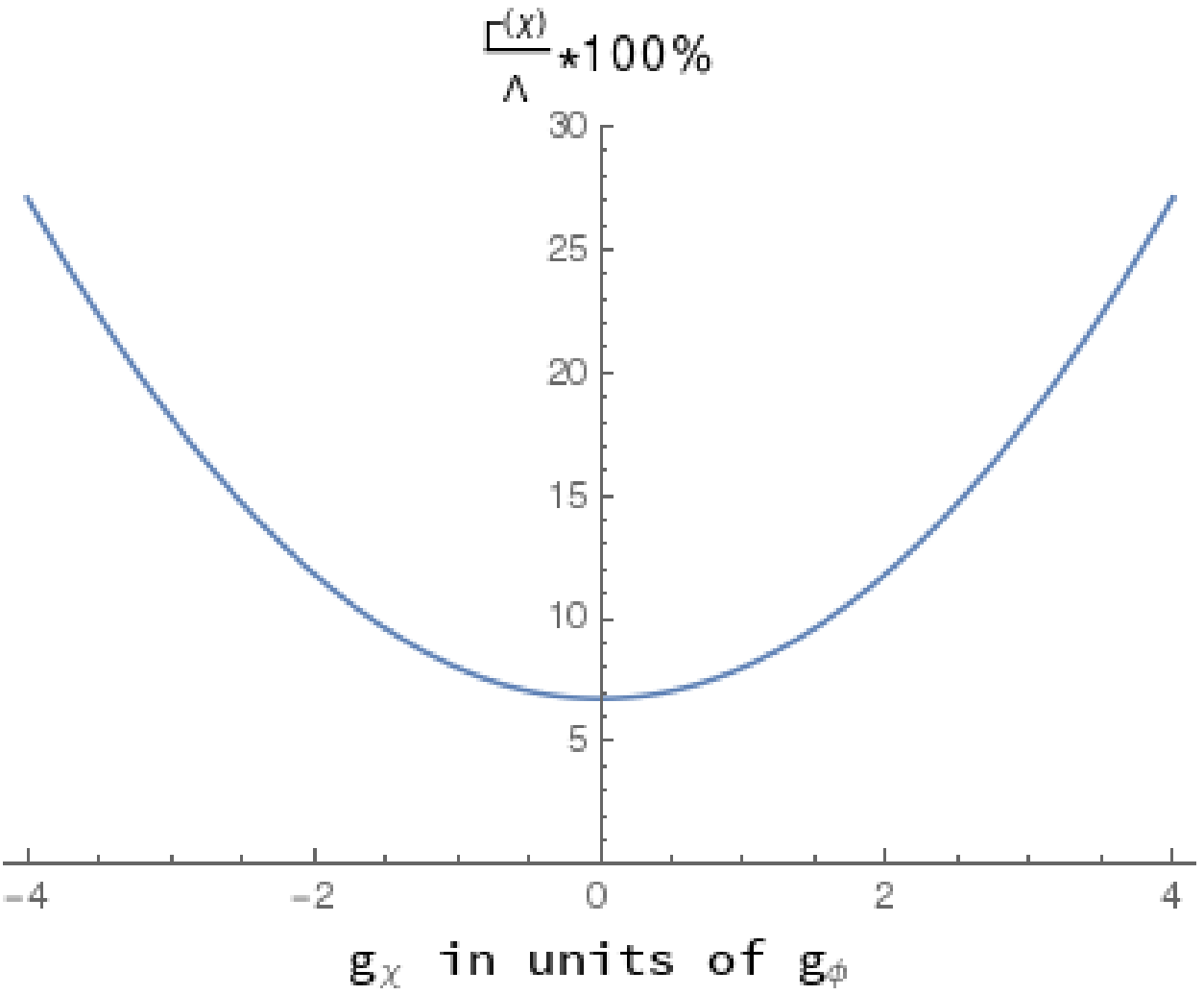}
    		\end{tabular}
    	\end{center}
    	\caption{Interference contribution estimation as a function of collision energy $\sqrt{p^2}$ and coupling $g_{\chi}$. Black dashed lines in the bottom of the contour plots show points where $p^2 = \mu^2$}
    	\label{fig:interference-contribution-comparison-gchi}
    \end{figure}

    To estimate the contribution of the interference term, we plot the quantity $\frac{|\sigma^{(interf.)}| - \sigma^{(\chi)}}{\sigma^{(\chi)}}$ in the contour plots for different values of $g_{\chi}$, $\Lambda$ and center-of-mass collision energy $\sqrt{p^2}$. The corresponding plots are shown in fig.\,\ref{fig:interference-contribution-comparison-gchi}. $\sqrt{p^2}$ is shown in these plots as a fraction of $\Lambda$, so a distance to the $\chi$ resonance is depicted. Negative values of $\frac{|\sigma^{(interf.)}| - \sigma^{(\chi)}}{\sigma^{(\chi)}}$ in the contour plots in fig.\,\ref{fig:interference-contribution-comparison-gchi} correspond to the  values of parameters for which $|\sigma^{(interf.)}| < \sigma^{(\chi)}$.
    Red dashed lines show parameters values when the contribution of $\chi$ boson to the total cross-section vanishes. In these points $\sigma^{(interf.)} + \sigma^{(\chi)} = 0$.

    As it follows from \eqref{cross-section-decomposition}, $\sigma^{(interf.)}\sim g_{\chi}^2$ and $\sigma^{(\chi)}\sim g_{\chi}^4$. Hence, in the limit $|g_{\chi}| \ll 1$, when the coupling $g_{\chi}$ between visible fermions and dark boson is small, the  interference term is much larger than the resonant one. As can be seen in fig.\,\ref{fig:interference-contribution-comparison-gchi}, in the vicinity of the axis $g_{\chi} = 0$  the ratio  $\frac{|\sigma^{(interf.)}| - \sigma^{(\chi)}}{\sigma^{(\chi)}}$ becomes very large.

    In the points where $|g_{\chi}| > |g_{\phi}|$ the interference term is larger than the resonant one only far from the $\chi$ resonance. According to \eqref{cross-section-decomposition} and the relations below, interference becomes weaker when the energy approaches the $\chi$ mass:
    \begin{equation*}
    \begin{split}
    & |\sigma^{(interf.)}|\sim g_{\chi}^2 (\Lambda^2 - p^2), \\
    & \sigma^{(\chi)} \sim g_{\chi}^4 (p^2 - \mu^2).
    \end{split}
    \end{equation*}
    Simultaneously, $\sigma^{(\chi)}$ rises faster than $\sigma^{(interf.)}$ when the coupling $g_{\chi}$ goes up.

    The mixing angle $\theta_{mix}$ as a function of the coupling $g_{\chi}$ is shown in the top two plots in fig.\,\ref{fig:interference-contribution-comparison-gchi}. As can be seen from these plots, the smaller values of the mixing angle correspond to the larger contribution of the interference term in the whole considered energy range. Indeed, according to \eqref{mixingAngleDef}, $\theta_{mix}\sim g_{\chi}\Lambda^{-2}$, so that $\theta_{mix}$ is small when either $g_{\chi}$ is feeble or $\Lambda^2 \gg \mu^2$, and $\chi$ resonance is located far.

    In our model, the mixing of  fields appears as a one-loop effect, and $\tan{2\theta_{mix}}\sim g_{\phi}g_{\chi}$. However, in  case of tree-level mixing, the corresponding angle of the diagonalization    may not depend on the tree level couplings. Thus, the interference and the resonant  contributions may be different from the ones depicted in fig.\,\ref{fig:interference-contribution-comparison-gchi}. For instance, the tree-level mixing appears if the model lagrangian contains the term $-\alpha^2\phi\chi$. Then the non-diagonal term in the  matrix element of the scattering will be proportional to $\alpha^2$, and therefore can not be neglected in the tree and one-loop approximations. Hence, in a general case of fields mixing the results in fig.\,\ref{fig:interference-contribution-comparison-gchi} do not hold.

    According to  modern experimental data, the couplings  between the visible and the dark matter should be small. Therefore in the contour plots in fig.\,\ref{fig:interference-contribution-comparison-gchi} we consider only the region where $|g_{\chi}| \ll 1$. Therein the interference term dominates over the resonance one in the all range of the discussed energies.

    Finally,  the contribution  due to relative interference does not depend on the value of the $\chi$ width $\Gamma$ and the coupling $G_{\chi}$, in the considered energy range.
    \section{Conclusions}
    In previous sections within the generalized Yukawa model we have analyzed the role of   particle coupling values  and masses  resulting in  the  creation of wide resonances in scattering processes. Such type resonances could  not be detected by the standard methods of direct searches for assuming the applicability of the NWA. To realize that, we divided the fields of the model in two classes -- "visible" and "dark" ones  -- and  considered  a number of  scenarios ensuring the DM boson gains a large resonance width in the invariant mass spectrum of final  states.  It turns out that the  limit of $3\%$  can be exceeded in many cases.  In the framework of our model, the conditions for that are the following:
    \begin{itemize}
    	\item DM particle is heavier than the visible one  -- $\Lambda>\mu$
    	\item mixing angle $|\theta_{mix}|\leq 10^{-5}$
    	\item interactions in the  visible sector are weaker than   that of  between the  dark  and visible particle  or between the  particles in the dark sector only.  That is,   if either $|g_{\chi}|\gg |g_{\phi}|$ or $|G_{\chi}|\gg |g_{\phi}|$. However, according to the modern experimental data the coupling between visible and dark matter has to be small. Thus, the first inequality should be ignored, and only the condition $|G_{\chi}| \gg |g_{\phi}|$ holds.
    \end{itemize}

    In certain cases, when the  width of dark boson $\geq 7\%$, the  NWA is not applicable to the resonance. Hence, such peaks are potentially invisible in direct  on-resonance searches. Moreover, if the  NWA is nevertheless used for  calculations of  the dark resonance contributions, it could result in the incorrect  estimate of the particle  mass and its couplings. Hence, an accurate  estimate of the maximal DM width is important for treating  experimental data. This does not depend on a specific model of the DM.

    In the present paper, the  s-channel process $\psi_1\bar{\psi}_1\rightarrow\psi_2\bar{\psi}_2$ was considered. It is similar to the process $l_1\bar{l}_1\rightarrow l_2\bar{l}_2$, where $l_i$ is the SM lepton of generation $i$. For example, this  can be the one $e^+e^-\rightarrow\mu^+\mu^-$. In this reaction, a hypothetical $Z^{\prime}$ boson could  appear as a virtual state, additionally to the SM photon and $Z$ boson. In this context, physical meaning of $\theta_{mix}$ limit is similar to that introduced in the phenomenology of $Z^{\prime}$ boson. This particle is presumably mixed with the SM $Z$ boson. The corresponding mixing angle $\theta_0$ is experimentally bounded to the  range $10^{-4}-10^{-3}$ \cite{bibl:gulov}. Simultaneously, $Z^{\prime}$ has to be much heavier than $Z$ \cite{bibl:pdg}. This corresponds to other condition derived above.  Hence we can conclude that the  parameters of visible particle resonance are independent of the characteristics  of the  dark sector. This is so until mixing between the visible and the  dark  bosons is small and  two  resonances  are located far enough one from another, avoiding interference between them. The presence of the upper limit on $\theta_{mix}$ is qualitatively important.

    The considered Yukawa model gave us a possibility for analyzing the role of the masses and couplings of particles. Other aspects of the problem such as group symmetry of the extended model and, hence, the content of the states we left behind it. In this approach we have obtained the set of conditions which should be taken into consideration when  searches for the DM particles are performed.
    As general conclusion, to overcome the problem of wide resonance states it is reasonable to additionally apply some non-resonant methods in order to detect these new states of matter.  Among them, the interference of dark and visible states should be taken into consideration at energies far from resonance peak. According to the analysis given above, the resonant term becomes larger than the interference contribution if and only if   the coupling to the dark particle is not extremely  small and the collision energy is of the same order as the $\chi$ resonance mass. If these conditions hold, on-resonance search methods could be used to identify the new particle. When the collision energy is much less than the mass of the new resonance, or coupling between the dark particle and a visible one is small, then interference term in a scattering process cross-section has to be taken into consideration. This term is dominant under these conditions, whereas the resonant term is suppressed. Hence, off-resonance search methods should be applied, to identify the new particle signal. The different type effective Lagrangians  could be derived and/or introduced to describe interactions between two worlds. These  problems  we left for the future.


    \section{Appendix A}
    Consider the one-loop matrix element of the process $\psi_1\bar{\psi}_1\rightarrow\psi_2\bar{\psi}_2$:
    \[\begin{split}
    &i\mathcal{M}(p_1;p_2\rightarrow p_3;p_4) = i\mathcal{M}^{(imp.\;Born)}(p_1;p_2\rightarrow p_3;p_4) + \\
    &+ i\mathcal{M}^{(box)}(p_1;p_2\rightarrow p_3;p_4).
    \end{split}\]
    In this equation $i\mathcal{M}^{(imp.\;Born)}$ stands for the improved Born approximation contribution, while $i\mathcal{M}^{(box)}$ comes from the box diagrams. Squared modulus of $i\mathcal{M}$ is given as follows:
    \[\begin{split}
    &|\mathcal{M}|^2 = |\mathcal{M}^{(imp.\;Born)}|^2 + \\
    &+ 2\Re\left\{\left(\mathcal{M}^{imp.\;Born}\right)^*\mathcal{M}^{(box)}\right\} + |\mathcal{M}^{(box)}|^2.
    \end{split}\]
    In the present research we assess the difference between $|\mathcal{M}|^2$ and $|\mathcal{M}^{(imp.\;Born)}|^2$, omitting term $|\mathcal{M}^{(box)}|^2$. We do so because of the technical difficulties behind $|\mathcal{M}^{(box)}|^2$ calculation. Hence, $|\mathcal{M}^{(imp.\;Born)}|^2$ is compared with the following approximation:
    \[|\mathcal{M}|^2\approx |\mathcal{M}^{(imp.\;Born)}|^2 + 2\Re\left\{\left(\mathcal{M}^{imp.\;Born}\right)^*\mathcal{M}^{(box)}\right\}.\]
    We found that for various ranges of the model parameters this $|\mathcal{M}|^2$ differs from $|\mathcal{M}^{(imp.\;Born)}|^2$ on the fraction less than $1\%$ of $|\mathcal{M}^{(imp.\;Born)}|^2$. Thus, $|\mathcal{M}^{(box)}|^2$ contribution would be even less, and the neglection of this term is valid. Eventually, we ignore the box diagrams contribution $\mathcal{M}^{(box)}$ as a whole, since its effect on particles' widths is negligibly small.

    \newpage
    \section{Appendix B}
    The error of the NWA approximation  is estimated as deviation of $\sigma_d$ from $\sigma_d^{(NWA)}$ as a fraction of $\sigma_d$. We consider such error as insignificant if its absolute value is less than or equal to corresponding deviation for visible resonance, described by contribution $\sigma_v$ and its approximation $\sigma_v^{(NWA)}$. The estimations were performed for the different model parameters fixed.
    \begin{table}[bh!]
    	\noindent\caption{$\Lambda$ variation}\tabcolsep4.5pt
    	
    	\noindent{
    		\begin{center}
    			\begin{tabular}{|c|c|c|c|}
    				\hline%
    				\multicolumn{1}{|c}{\rule{0pt}{5mm}$\rho$, \%}%
    				& \multicolumn{1}{|c}{$\Lambda$}
    				& \multicolumn{1}{|c}{$\frac{|\sigma_v-\sigma_v^{(\text{NWA})}|}{\sigma_v}$, \%}
    				& \multicolumn{1}{|c|}{$\frac{|\sigma_d-\sigma_d^{(\text{NWA})}|}{\sigma_d}$, \%}\\[2mm]%
    				\hline%
    				\rule{0pt}{5mm}1.26 & 0.25$\mu $ & 4.839 & 2.077 \\
    				6.3 & 0.9$\mu $ & 4.955 & 5.63 \\
    				7.867 & 3$\mu $ & 4.232 & 5.898 \\
    				7.983 & 6$\mu $ & 3.468 & 5.921 \\[2mm]%
    				\hline
    			\end{tabular}
    		\end{center}
    	}%
    	\label{table:lambda-var}
    \end{table}
    \begin{table}[bh!]
    	\noindent\caption{$M$ variation}\tabcolsep4.5pt
    	
    	\noindent{
    		\begin{center}
    			\begin{tabular}{|c|c|c|c|}
    				\hline%
    				\multicolumn{1}{|c}{\rule{0pt}{5mm}$\rho$, \%}%
    				& \multicolumn{1}{|c}{M}
    				& \multicolumn{1}{|c}{$\frac{|\sigma_v-\sigma_v^{(\text{NWA})}|}{\sigma_v}$, \%}
    				& \multicolumn{1}{|c|}{$\frac{|\sigma_d-\sigma_d^{(\text{NWA})}|}{\sigma_d}$, \%}\\[2mm]%
    				\hline%
    				\rule{0pt}{5mm}8.033 & 0 & 3.468 & 6.004 \\
    				8.1 & $m_1$ & 3.468 & 7.703 \\
    				7.983 & $\approx$ 2000 $m_1$ & 3.468 & 5.921 \\[2mm]%
    				\hline
    			\end{tabular}
    		\end{center}
    	}%
    	\label{table:darkm-var}
    \end{table}
    \begin{table}[bh!]
    	\noindent\caption{$g_{\chi}$ variation}\tabcolsep4.5pt
    	
    	\noindent{
    		\begin{center}
    			\begin{tabular}{|c|c|c|c|}
    				\hline%
    				\multicolumn{1}{|c}{\rule{0pt}{5mm}$\rho$, \%}%
    				& \multicolumn{1}{|c}{$g_{\chi}$}
    				& \multicolumn{1}{|c}{$\frac{|\sigma_v-\sigma_v^{(\text{NWA})}|}{\sigma_v}$, \%}
    				& \multicolumn{1}{|c|}{$\frac{|\sigma_d-\sigma_d^{(\text{NWA})}|}{\sigma_d}$, \%}\\[2mm]%
    				\hline%
    				\rule{0pt}{5mm}7.975 & $1.g_{\phi}$ & 3.468 & 5.921 \\
    				13.81 & $2.438g_{\phi}$ & 3.678 & 9.283 \\
    				19.41 & $3.25g_{\phi}$ & 3.856 & 14.81 \\
    				26.23 & $4.g_{\phi}$ & 3.945 & 22.38 \\
    				56.95 & $5.9g_{\phi}$ & 3.353 & 51.94 \\[2mm]%
    				\hline
    			\end{tabular}
    		\end{center}
    	}%
    	\label{table:gchi-var}
    \end{table}
    \begin{table}[bh!]
    	\noindent\caption{$G_{\chi}$ variation}\tabcolsep4.5pt
    	
    	\noindent{
    		\begin{center}
    			\begin{tabular}{|c|c|c|c|}
    				\hline%
    				\multicolumn{1}{|c}{\rule{0pt}{5mm}$\rho$, \%}%
    				& \multicolumn{1}{|c}{$G_{\chi}$}
    				& \multicolumn{1}{|c}{$\frac{|\sigma_v-\sigma_v^{(\text{NWA})}|}{\sigma_v}$, \%}
    				& \multicolumn{1}{|c|}{$\frac{|\sigma_d-\sigma_d^{(\text{NWA})}|}{\sigma_d}$, \%}\\[2mm]%
    				\hline%
    				\rule{0pt}{5mm}1.231 & $0.01g_{\phi}$ & 3.482 & 2.429 \\
    				1.378 & $0.5g_{\phi}$ & 3.482 & 1.59 \\
    				3.765 & $2.g_{\phi}$ & 3.483 & 3.259 \\
    				7.99 & $3.25g_{\phi}$ & 3.484 & 5.955 \\
    				17.78 & $5.g_{\phi}$ & 3.485 & 11.26 \\
    				40.96 & $7.5g_{\phi}$ & 3.489 & 15.91 \\[2mm]%
    				\hline
    			\end{tabular}
    		\end{center}
    	}%
    	\label{table:dark-gchi-var}
    \end{table}

\end{document}